\documentclass[12pt,twoside]{article}

\pdfoutput=1

\usepackage[usenames]{color}
\usepackage{lscape}
\usepackage{amssymb}
\usepackage{rotating}
\usepackage[T1]{fontenc}
\usepackage{graphicx}
\usepackage{fancyhdr}
\usepackage{amsmath}
\usepackage{ae}
\usepackage{aecompl}
\usepackage{hyperref}

\def\pa{\partial}
\def\nn{\nonumber \\}
\def\ov{\overline}

\def\de{{\rm d}}

\newcommand{\be}{\begin{equation}}
\newcommand{\ee}{\end{equation}}
\newcommand{\ba}{\begin{eqnarray}}
\newcommand{\ea}{\end{eqnarray}}

\newlength{\dinwidth}
\newlength{\dinmargin}
\begin{document}

\thispagestyle{empty}
\begin{flushright}
Cavendish--HEP--12/09\\
DAMTP--2012--35\\
CERN-PH-TH-2012-108\\
CPHT-RR 017.0512 \\

\end{flushright}

\vspace*{1cm}

\centerline{\Large\bf Inverted sfermion mass hierarchy}
\vspace*{2mm}
\centerline{\Large\bf and the Higgs boson mass in the MSSM}

\vspace*{5mm}

\vspace*{5mm} \noindent
\vskip 0.5cm
\centerline{\bf
Marcin Badziak${}^{a,b,}$\footnote[1]{M.Badziak@damtp.cam.ac.uk},
Emilian Dudas${}^{c,d,}$\footnote[2]{Emilian.Dudas@cpht.polytechnique.fr}
Marek Olechowski${}^{e,}$\footnote[3]{Marek.Olechowski@fuw.edu.pl},
Stefan Pokorski${}^{e,f,}$\footnote[4]{Stefan.Pokorski@fuw.edu.pl}
}
\vskip 5mm
\begin{center}
{${}^a$\em Department of Applied Mathematics and Theoretical Physics,
Centre for Mathematical Sciences, University of Cambridge,
Wilberforce Road, Cambridge CB3 0WA, United Kingdom \\[6pt]

${}^b$Cavendish Laboratory, University of Cambridge, J.J. Thomson Avenue,
\\ Cambridge CB3 0HE, United Kingdom \\[6pt]

${}^c$ CPhT, Ecole Polytechnique, 91128 Palaiseau, France \\[6pt]

${}^d$  Department of Physics, Theory Division, CH-1211,Geneva 23, Switzerland
\\[6pt] 

${}^e$ Institute of Theoretical Physics, Faculty of Physics,
University of Warsaw\\
ul.\ Ho\.za 69, PL--00--681 Warsaw, Poland\\[6pt]

${}^f$ TUM-IAS, Technische Universitat Munchen, Lichtenbergstr. 2A, D-85748 Garching, Germany}
\end{center}

\vskip 1cm

\centerline{\bf Abstract}
It is shown that MSSM with first two generations of squarks and sleptons much heavier than the third one naturally predicts the maximal stop mixing as a consequence of the RG evolution, with vanishing (or small) trilinear coupling at the high scale. The Higgs boson is generically heavy, in the vicinity of 125 GeV. In this inverted hierarchy scenario, motivated by the supersymmetric FCNC problem and models for fermion masses based on horizontal symmetries,
the mass of the lightest stop is $\mathcal{ O}(0.5)$ TeV  and  of gluino - $\mathcal{O}(2-3)$ TeV. The LSP can be either higgsino or bino or a mixture of both and it can be a good dark matter candidate.

\vskip 3mm

\newpage

\section{Introduction}

The issue of naturalness of the Higgs potential has for long been a driving principle
for theoretical ideas going beyond the Standard Model. This is a qualitative and only
a theoretical argument. However, if one abandons it, there is no reason to expect new
physics at the LHC. If one takes it too literally, new particles should have already been
observed at the LHC or even at LEP.

In supersymmetric (SUSY) models, MSSM in particular, it has been appreciated from the beginning
of the supersymmetric phenomenology  that only part of the superpartner spectrum is concerned
 by the naturalness argument, namely higgsinos, the third generation sfermion masses and gluinos
(and to a much lesser extent other gauginos). Those are the particles whose masses enter the Higgs
potential either at the tree level (higgsinos) or via loop effects, enhanced by large couplings.
Naturalness argument suggests that those particles are rather light but says nothing about the
masses of the first  and the second generation sfermions.

On the other hand, it has very early been observed that the supersymmetric FCNC and CP violation
problems can be substantially eased the heavier the first-two-generation sfermions are. Thus the concept
of naturalness and the FCNC supersymmetric problem lead together to the expectation of a split sfermion
spectrum. Furthermore, in fermion mass models based on horizontal symmetries, the sfermion mass spectrum
has been linked to the hierarchical fermion masses. The predicted  pattern is the so-called inverted
hierarchy (IH) of the sfermion masses.

In this paper we investigate the predictions for the lightest Higgs boson mass in the MSSM with inverted
hierarchy of  sfermion masses. Due to 2-loop effects, heavy first two generations play significant role in
reaching the Higgs mass in the region of 125 GeV which seems to be favoured by the recent ATLAS \cite{atlas_higgs} and CMS \cite{cms_higgs} data. This is because the maximal stop mixing is obtained
from the RG evolution effects, with initial $A_0\leq m_0(3)$ and even with $A_0=0$, with $m_0(3)$ being  the scale of the 3rd generation sfermion masses. The lightest stop
is predicted to be in the 500-1000 GeV mass range  and gluinos are 2-3 TeV heavy.
The model is less fine-tuned than CMSSM.

Various phenomenological aspects of the inverted hierarchy of the sfermion
masses and its impact on the Higgs mass have also been discussed in a recent
paper \cite{Baer_NatSUSY}. Our findings are in a qualitative agreement with
the results of ref.\ \cite{Baer_NatSUSY}. Our predictions for the Higgs boson
mass are often a couple of GeV higher and, as we discuss in more detail at the
end of Section \ref{sec:IH}, this can be traced to the 1-loop
\cite{Baer_NatSUSY} versus 2-loop (in our case) calculation of the Higgs mass
and to the not fully overlapping range of the investigated parameter space.

The paper is organized as follows. In section (\ref{sec:higgs_low}) we discuss the implications of the low energy MSSM spectrum on the Higgs boson mass. A crucial role of the stop mixing in reaching large values of the Higgs boson mass is emphasized. In section (\ref{sec:IH}) we perform a detailed study of the predictions for the Higgs boson mass in the IH scenario. We also review there some of the theoretical ideas leading to the IH of sfermion masses. In section \ref{sec:pheno} other phenomenological constraints and implications for the LHC are discussed. Several benchmark points are presented. Our conclusions are presented in section \ref{sec:concl}.

\section{The Higgs boson mass in the electroweak scale MSSM}
\label{sec:higgs_low}

It is well known that at the tree level the mass of the lightest Higgs boson
in the MSSM is bounded from above by $M_Z|\cos 2\beta|$.
However, the tree-level upper bound on the Higgs boson
mass is uplifted at the quantum level when the effects of (soft) SUSY breaking
are taken into account. The magnitude of the loop corrections depends mainly on the properties of the stop sector. In the
decoupling limit, $m_A\gg M_Z$ where $A$ denotes the CP-odd scalar, the Higgs boson mass corrected by the dominant
one-loop contribution is given by \cite{HHH}:
\begin{equation}
\label{mh_1loop}
 m_h^2\approx M_Z^2\cos^2 2\beta + \frac{3g^2m_t^4}{8 \pi^2 m_W^2}
\left[\ln\left(\frac{M_{\rm SUSY}^2}{m_t^2}\right)+\frac{X_t^2}{M_{\rm SUSY}^2}
\left(1-\frac{X_t^2}{12M_{\rm SUSY}^2}\right)\right] \,,
\end{equation}
where $M_{\rm SUSY}\equiv\sqrt{m_{\tilde{t}_1}m_{\tilde{t}_2}}$ ($m_{\tilde{t}_i}$
are the eigenvalues of the stop mass matrix  at $M_{\rm SUSY}$ in the $\ov{DR}$ renormalization scheme) and $X_t\equiv
A_t-\mu/\tan\beta$ with $A_t$ being SUSY breaking top trilinear coupling at $M_{\rm SUSY}$.
It follows from the above formula that sizable corrections to the Higgs
boson mass can be obtained when the stops are substantially heavier than the
top quark. Since the contribution to the Higgs boson mass from stop mixing may
be significant, the precise values of the stop masses required to obtain a
given value of the Higgs boson mass are quite sensitive to the ratio
$X_t/M_{\rm SUSY}$. It follows from eq.~(\ref{mh_1loop}) that the maximal
contribution from stop mixing is obtained for $|X_t|/M_{\rm SUSY}= \sqrt{6}$.

Equation (\ref{mh_1loop}) was derived under the assumption of a relatively small
mass splitting between the stops. The generalization of this formula to the
case of a large splitting can be found e.g.\ in
Ref.~\cite{HHH}. One can then show that,
for fixed $M_{\rm SUSY}$, the value of $|X_t|/M_{\rm SUSY}$ giving
the maximal correction from stop mixing to the Higgs mass, as well
as the value of this maximal correction increase with the splitting
between the two eigenstates of the running stop mass matrix.

There are theoretical uncertainties
in the prediction of the Higgs boson mass. They originate from
unknown higher order corrections and from the limited
experimental knowledge of the SM parameters, mainly the top mass and the
strong gauge coupling constant. The estimated uncertainty is about 3 GeV \cite{Allanach_higgs}.\footnote{The dominant 3-loop contribution to the Higgs mass
has been calculated in \cite{Higgs_3loop}. In the case of CMSSM it is positive and of the order of 1-3 GeV with precise value depending on the soft terms \cite{Kant}.}
Taking
this into account the Higgs boson masses in the range 122-128 GeV are
consistent with the 125 GeV Higgs interpretation of the excesses observed by
the LHC experiments.
In the present paper we use $m_t^{\rm pole}=173.3$
GeV and $\alpha_s(M_Z)=0.1187$.

In our numerical studies we use a modified version of SOFTSUSY v3.2.4
\cite{softsusy} which employs the two-loop formulae for the Higgs boson mass
\cite{Slavich}. The original version of SOFTSUSY refuses to calculate the
spectrum if there are any negative running ($\ov{DR}$) squared masses at the
$M_Z$ scale, even if there are no tachyons at the $M_{\rm SUSY}$ scale where the
Higgs potential is minimized. The reason for this is that in SOFTSUSY the
running parameters are used to compute the SUSY corrections to the SM
parameters (such as gauge or Yukawa couplings) at the $M_Z$ scale. Such a
procedure leads to excluding some of the parameter space which is perfectly
viable from the theoretical point of view. Since this part of the parameter
space is important for the studies in the present paper we have modified
SOFTSUSY in such a way that the pole masses are used in the calculation
of the SUSY corrections to the SM parameters and the tachyons are signalized
only if there are any negative squared running masses at the $M_{\rm SUSY}$ scale.

\begin{figure}[t!]
  \begin{center}
    \includegraphics[width=0.65\textwidth]{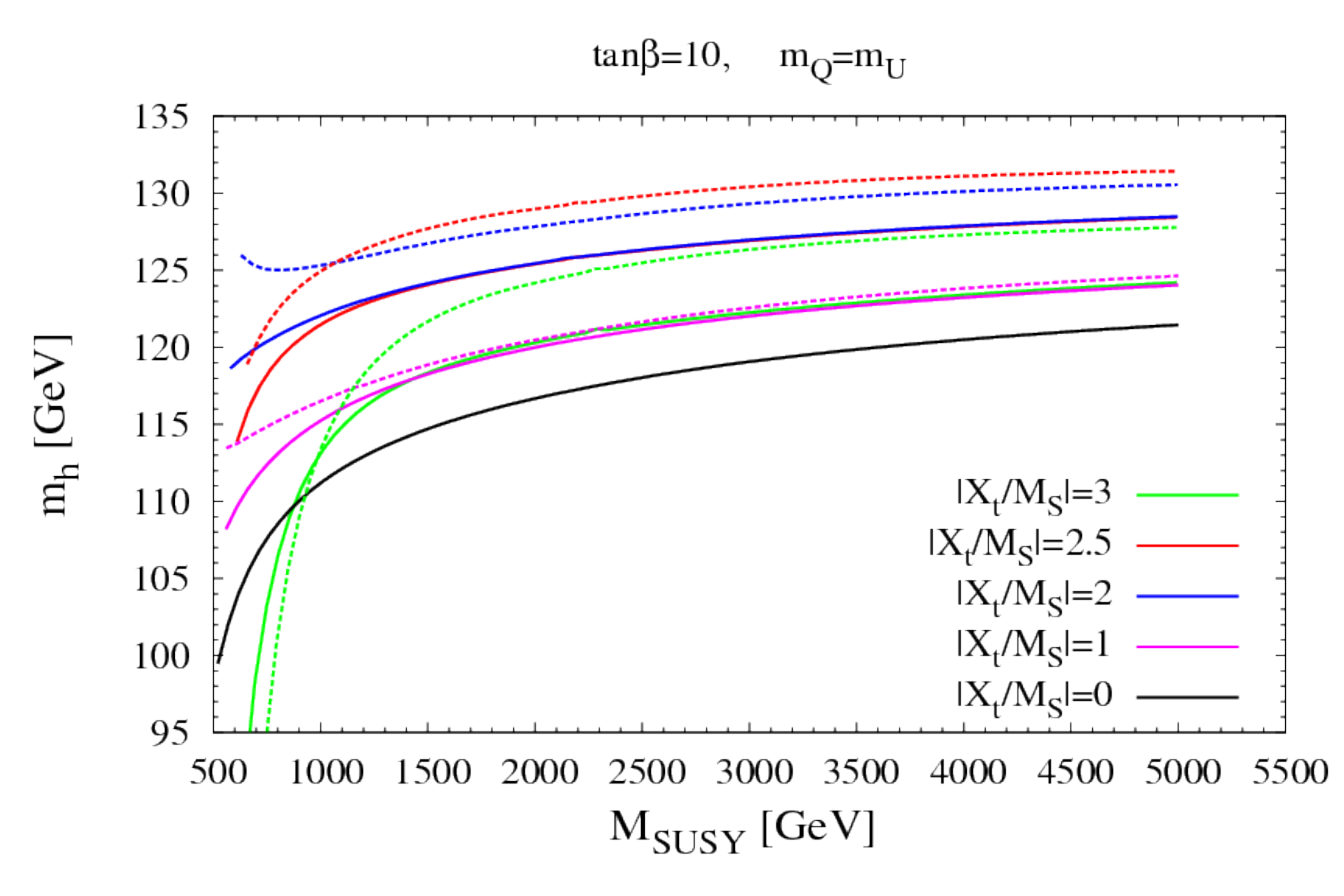}
    \includegraphics[width=0.65\textwidth]{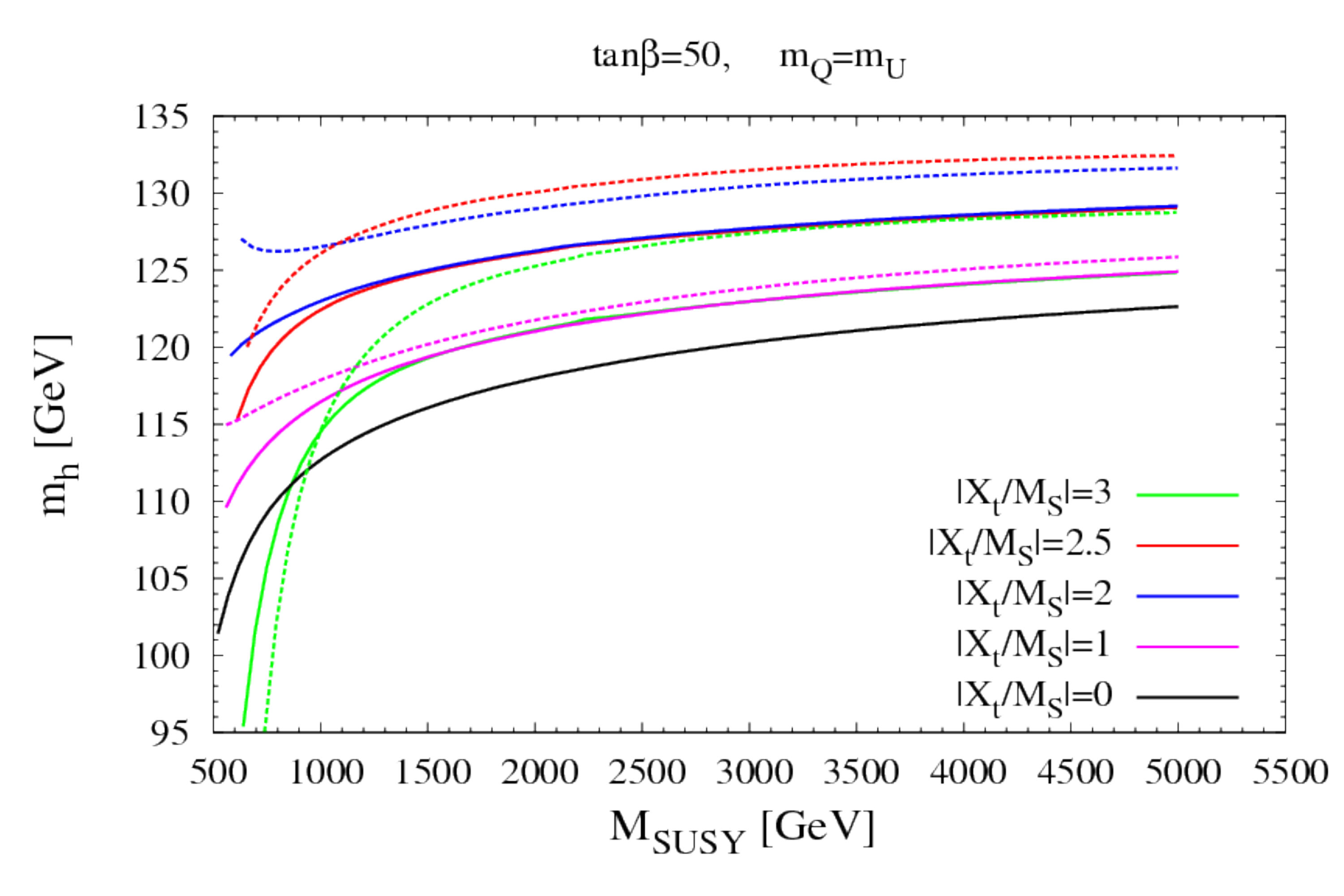}
    \caption{The Higgs boson mass versus $M_{\rm SUSY}$ for various values of
      $X_t/M_{\rm S}$ and $\tan\beta=10$ (top) or $\tan\beta=50$ (bottom). The
      solid (dashed) lines correspond to the negative (positive) values of
      $X_t$. All the other MSSM parameters (defined at the $M_{\rm SUSY}$ scale)
      are fixed to be 2 TeV except for $M_2=\mu=m_A=1$ TeV, $M_1=0$ (to ensure
      that a neutralino is the LSP),  $A_b=A_{\tau}=0$. Analogous plots for
      negative $\mu=-1$ TeV are very similar.}
    \label{fig:mh_low_msusy}
  \end{center}
\end{figure}

In Figure \ref{fig:mh_low_msusy} we plot the Higgs boson mass versus
$M_{\rm SUSY}$ for various values of $X_t/M_{\rm S}$, where $M_{\rm S}\equiv\sqrt{m_Q m_U}$
and $m_{Q(U)}$ are the running soft masses of the third generation left-handed
squark doublet (right-handed up-type squark) at the $M_{\rm SUSY}$ 
scale.\footnote{
The parameter $M_{\rm S}$ is defined using the soft masses while $M_{\rm SUSY}$ introduced
in Eq.~(\ref{mh_1loop}) is defined using the eigenvalues of the running stop
mass matrix. The values of $M_{\rm S}$ and $M_{\rm SUSY}$ are quite similar in most part
of parameter space. The relative difference between $M_{\rm S}$ and $M_{\rm SUSY}$
increases as $M_{\rm S}$ decreases and in some cases may reach 10\%.
In some figures we use $M_{\rm S}$ for technical reasons. In SOFTSUSY it is
possible to fix the value of $M_{\rm S}$ while $M_{\rm SUSY}$ is obtained as a result
of an iteration procedure.
}
The data used in this plot were obtained by scanning $M_{\rm S}=m_Q=m_U$ up to 5 TeV
while keeping fixed all other MSSM parameters defined at the $M_{\rm SUSY}$
scale. 
It is clear from this figure that the scenario with  vanishing or very small
stop mixing is incompatible with the Higgs boson mass of about 125 GeV
(suggested by recent ATLAS \cite{atlas_higgs} and CMS \cite{cms_higgs} results).
The Higgs boson mass of 120 GeV may be obtained in the absence of the
mixing in the stop sector but only for very heavy stops.
On the other hand, when the contribution to the
Higgs boson mass from the stop mixing is large, stop masses of about
1 TeV may be consistent with the Higgs mass of 125 GeV.
The largest Higgs mass is obtained for $|X_t/M_{\rm S}|$ between 2 and 2.5,
in agreement with the leading one-loop formula (\ref{mh_1loop}).
Notice also that even though the expression (\ref{mh_1loop}) does not
depend on the sign of $X_t$, for the positive $X_t$ the Higgs mass
is enhanced more than for the negative $X_t$. This asymmetry arises
due to SUSY threshold effects on the top Yukawa
coupling which depend on the product of the gluino mass and $A_t$
\cite{Xtasymmetry}.

\begin{figure}[t!]
  \begin{center}
    \includegraphics[width=0.65\textwidth]{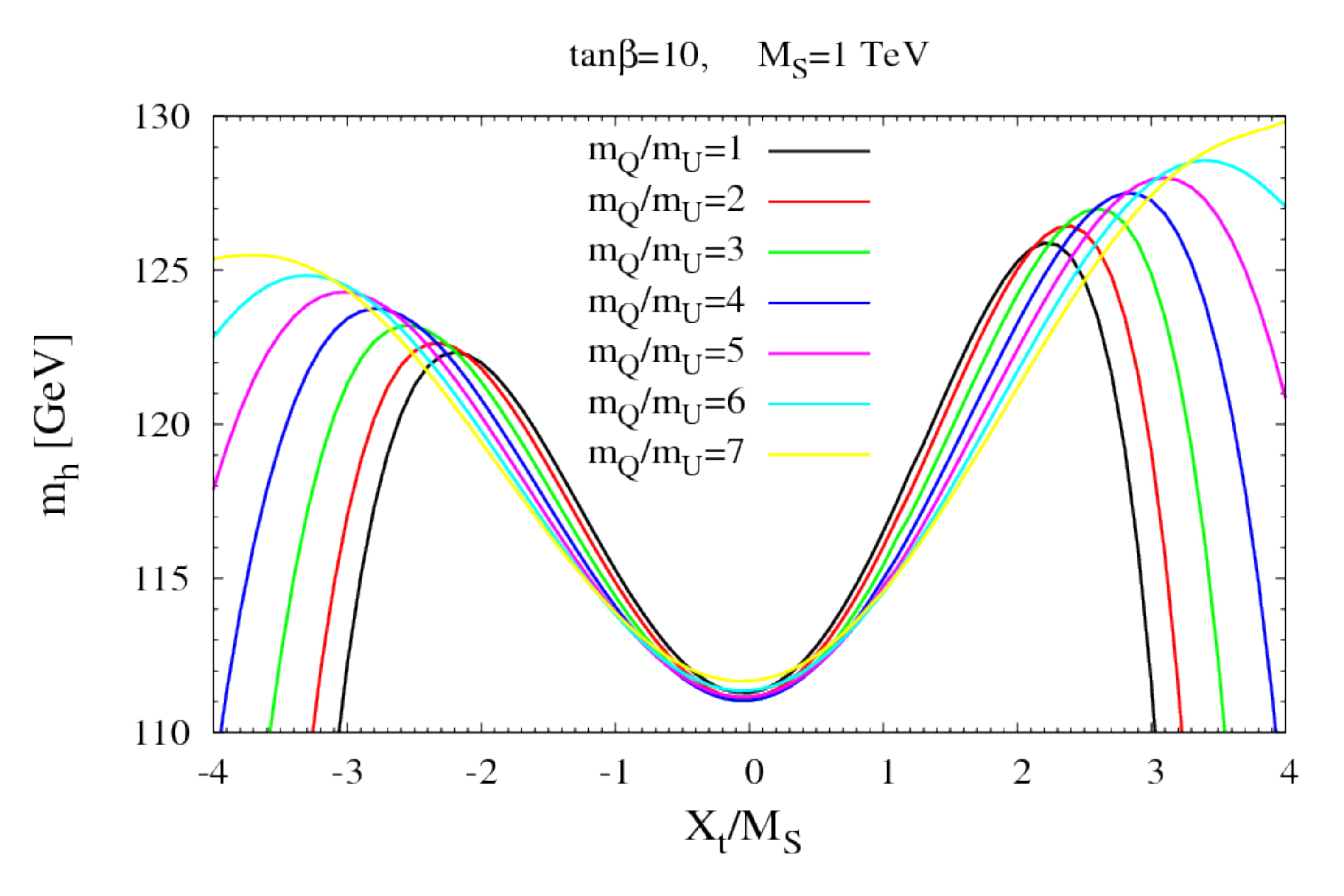}
    \caption{The Higgs boson mass versus $X_t/M_{\rm S}$ for various values of
      the ratio $m_Q/m_U$ keeping fixed $M_{\rm S}=1$ TeV assuming
      $\tan\beta=10$. All other MSSM parameters at $M_{\rm SUSY}$ 
      are the same as in the case of Figure \ref{fig:mh_low_msusy}.  }
    \label{fig:mh_xt_low}
  \end{center}
\end{figure}

The dependence of the Higgs mass on the stop masses splitting
is shown in Figure \ref{fig:mh_xt_low}. The Higgs mass versus $X_t/M_{\rm S}$
is plotted there for various values of the ratio $m_Q/m_U$ keeping
fixed the value of $M_{\rm S}=1$ TeV. While for $m_Q=m_U$ the local maxima of the
Higgs mass occur at $|X_t/M_{\rm S}|_{\rm max}\approx 2.2$ the positions of these
maxima move towards the larger values of $|X_t/M_{\rm S}|_{\rm max}$ exceeding 3 for
$m_Q/m_U\gtrsim 5$. Moreover, the value of the Higgs mass corresponding to
$|X_t/M_{\rm S}|_{\rm max}$ in the case of large stop mass splitting may be larger
by several GeV as compared to the unsplit case. Figure \ref{fig:mh_xt_low} shows that the bigger stop mass splitting is the
bigger maximal Higgs mass may be obtained.
It is also interesting to note
that for a given $|X_t/M_{\rm S}|\lesssim 2$ the mass splitting between the stops
tends to suppress the Higgs mass.

For large values of $\tan\beta$ for which the bottom and tau Yukawa couplings
are of the same order as the top Yukawa coupling the loop corrections from the
sbottom and stau sectors may also be significant. These corrections tend to
reduce the Higgs boson mass and become quite large for $|\mu|\gg m_{\tilde{b}}$
and/or $|\mu|\gg m_{\tilde{\tau}}$, especially when $\mu M_3<0$ since this
typically leads to the enhancement of the bottom Yukawa coupling due to SUSY
threshold corrections important for large $\tan\beta$.
These effects have been recently discussed in \cite{Carena_gamma}.
However, such reduction of the Higgs mass at large $\tan\beta$ is not present
in Figure \ref{fig:mh_low_msusy} 
(also for Figure \ref{fig:mh_xt_low} the (not shown) plots for 
large $\tan\beta$ are very similar to those for $\tan\beta=10$)
which is a consequence of the fact that $|\mu|$ is chosen there to be smaller than the sbottom and the
stau masses. For the same reason the results for negative $\mu$ are
very similar to those for positive $\mu$ presented in these Figures.
As a matter of fact, for our choice of the parameters the Higgs mass for
$\tan\beta=50$ is slightly larger than for $\tan\beta=10$. This
can be attributed to the tree-level contribution which grows with
$\tan\beta$.

\section{The Higgs boson mass in  IH  Model}
\label{sec:IH}

As discussed in the previous section, $M_{\rm SUSY}\gtrsim 1$ TeV and  a big contribution from the stop mixing
to the Higgs mass is necessary to obtain $m_{h}\gtrsim 122$ GeV. The soft
trilinear term $A_t$ at the EW scale should not be much different from its
optimal value corresponding to $|X_t/M_{\rm SUSY}|_{\rm max}$ which maximizes the
Higgs mass (for fixed other parameters). Typically, such optimal value of
$|A_t|$ is 2 to 3 times bigger than $M_{\rm SUSY}$.
In  UV models,  the values of $M_{\rm SUSY}$ and the stop mixing at the EW scale depend on the  pattern of soft terms fixed at a (usually  high) scale of supersymmetry breaking mediation  and on the RG evolution.  In this paper
we are interested in the prediction for the lightest Higgs boson mass in the scenario with the inverted hierarchy
of squark masses, motivated by the naturalness arguments and by the flavour models based on horizontal symmetries. We shall not refer to any specific model. The Higgs boson mass is calculated for a
given set of the boundary conditions for the soft terms at the GUT scale.

Since the large stop mixing at the EW scale is crucial for the successful prediction for the Higgs boson  mass,
before going to the IH scenario,   it is useful to discuss when large values of $|A_t/M_{\rm SUSY}|$ at the EW scale  can be obtained.
Due to the RGE running, the low scale value of $A_t$ is a linear combination
of the high energy scale gaugino masses and trilinear terms. In the case of
the universal high energy scale boundary conditions for these parameters
($M_{1/2}$ and $A_0$, respectively), the EW scale value of $A_t$ is
approximately given by
\begin{equation}
A_t \approx - 1.6 M_{1/2} + 0.35 A_0
\,.
\label{Atcoeff}
\end{equation}
The precise values of the coefficients in the above equation depend on the
gauge and Yukawa couplings, so implicitly (among other parameters) on the
values of the top mass and $\tan\beta$. The coefficient in front of $A_0$ is
relatively small because the top quark is heavy. It goes to zero when the top
quark mass approaches its infra-red quasi-fixed point value
\cite{Carena_lowtanb,Carena}.

Equation (\ref{Atcoeff}) is valid at the 2-loop RGE level. For the sake of definiteness we give in this paper all the
coefficients obtained from the RG running (unless stated otherwise)  computed at the scale $Q=1.5$ TeV and the gauge and Yukawa coupling at the GUT scale consistent with radiative electroweak symmetry breaking (REWSB) for $\tan\beta=10$, $M_{1/2}=700$ GeV, $A_0=-3$ TeV, the scalar mass of the third generation $m_0(3)=3$ TeV and that of the first and second generation $m_0(1,2)=10$ TeV ($A_t$ in eq.\ (\ref{Atcoeff}) does not depend on the latter).
In most of the paper, unless otherwise stated, we also take 
universal Higgs soft masses  at the GUT scale 
$m_0({H_u}) = m_0({H_d})= m_0(3)$.

The values of $m_Q$ and $m_U$ at the EW scale in terms of the GUT scale values of the soft terms are given by:
\begin{align}
\label{mQcoeff}
 m_Q^2\approx 3.1M_{1/2}^2+0.1A_0M_{1/2}-0.04A_0^2+0.65m_0(3)^2-0.03m_0(1,2)^2 \,, \\
\label{mUcoeff}
 m_U^2\approx 2.3M_{1/2}^2+0.2A_0M_{1/2}-0.07A_0^2+0.35m_0(3)^2-0.02m_0(1,2)^2 \,.
\end{align}
The dependence of the 3rd generation squark masses at the EW scale  on the $m_0(1,2) $ is a 2-loop effect and it is negligible for small values of $m_0(1,2)$.  However, as was pointed out in \cite{AHM}, for very large scalar masses of the first two generations it may lead to tachyonic stops  due to the (small) negative coefficient in front of $m_0(1,2)^2$ in the expressions (\ref{mQcoeff})-(\ref{mUcoeff}).

An inspection of Eqs.~(\ref{Atcoeff})-(\ref{mUcoeff}) shows that, as long as the 2-loop effect is negligible, e.g.\ in CMSSM, "maximal"  stop mixing can be obtained only for   sufficiently large values of $A_0$.  The needed value of
$A_0$ depends on the relative magnitude of $M_{1/2}$ and $m_0$. For $m_0\gg
M_{1/2}$,  $A_0/m_0\approx \pm2$ is required, with negative sign more
effective for not too large a ratio $m_0/M_{1/2}$. For $ m_0\ll M_{1/2}$ one
needs $A_0/M_{1/2}\approx -3.5$  and for  $M_{1/2}\approx m_0$ that ratio has
to be about $-4$. These conclusions agree with a recent study \cite{Brummer}.
The reason a rather large $|A_0|$ is needed  is that the RG evolution of the
dividend and of the divisor in the ratio
$A_t^2/(m_Q m_U)$ is correlated (leading to $A_t^2/M_{\rm SUSY}^2\lesssim 1$ for small $|A_0|$).  In consequence, a 125 GeV Higgs boson requires a heavy spectrum and significant cancellations in the Higgs potential.

\begin{figure}[t]
  \begin{center}
    \includegraphics[width=\textwidth, height=0.34\textheight]{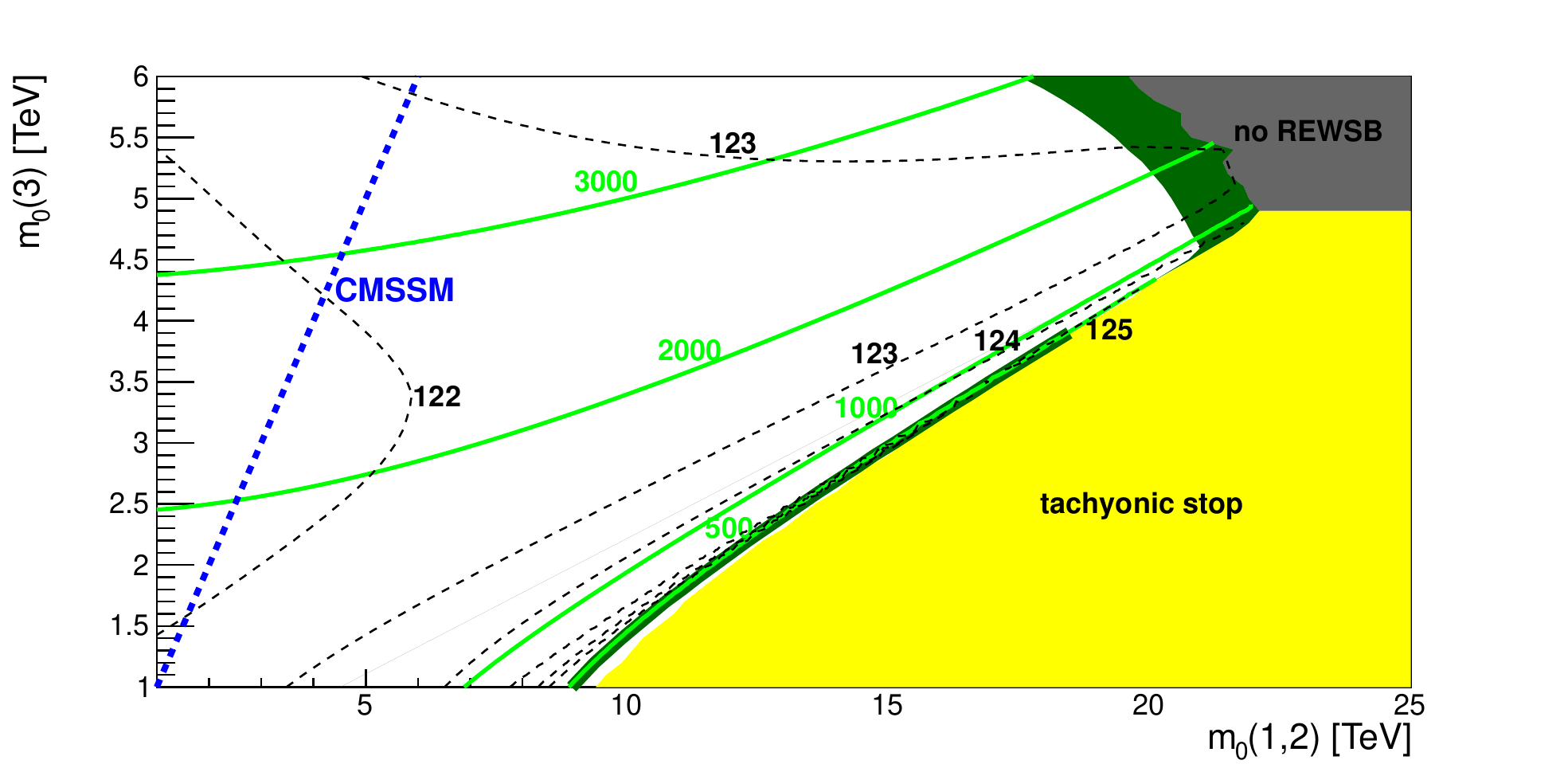}
    \caption{Contours of the Higgs boson mass (black dashed line), the lighter stop mass (solid green line)  for $\mu>0$, $\tan\beta=10$, $M_{1/2}=1$ TeV and $A_0=-2$ TeV. The yellow ``tachyonic stop'' and the grey ``no REWSB'' ($\mu^2<0$) regions  are excluded. In the dark green region the relic density of neutralinos gives $\Omega_{\rm DM}h^2<0.1288$ \cite{WMAP7}.  }
    \label{fig:mh_IMH_A0_2000}
  \end{center}
\end{figure}
\begin{figure}[p]
  \begin{center}
    \includegraphics[width=\textwidth, height=0.31\textheight]{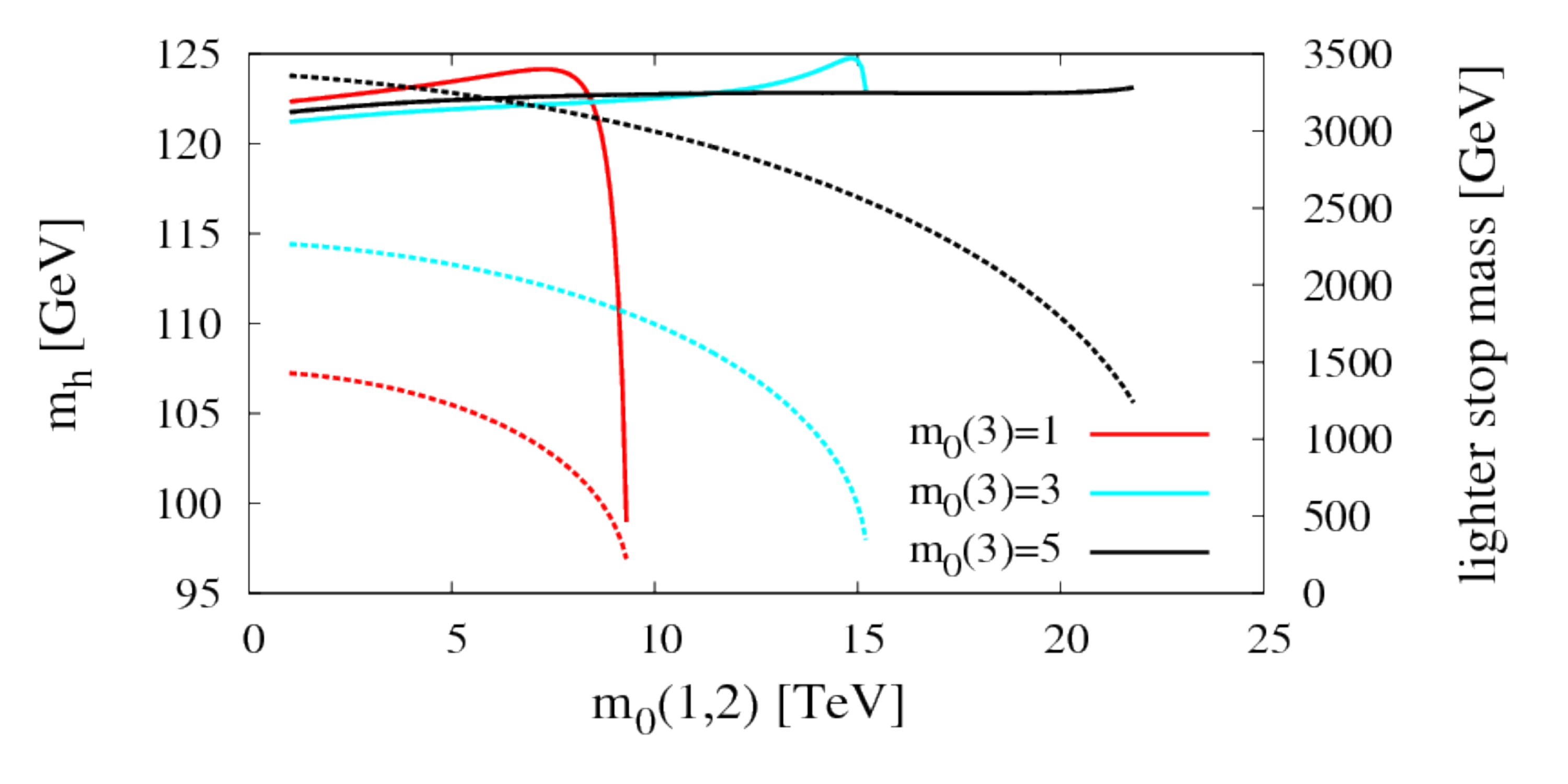}
    \includegraphics[width=\textwidth, height=0.31\textheight]{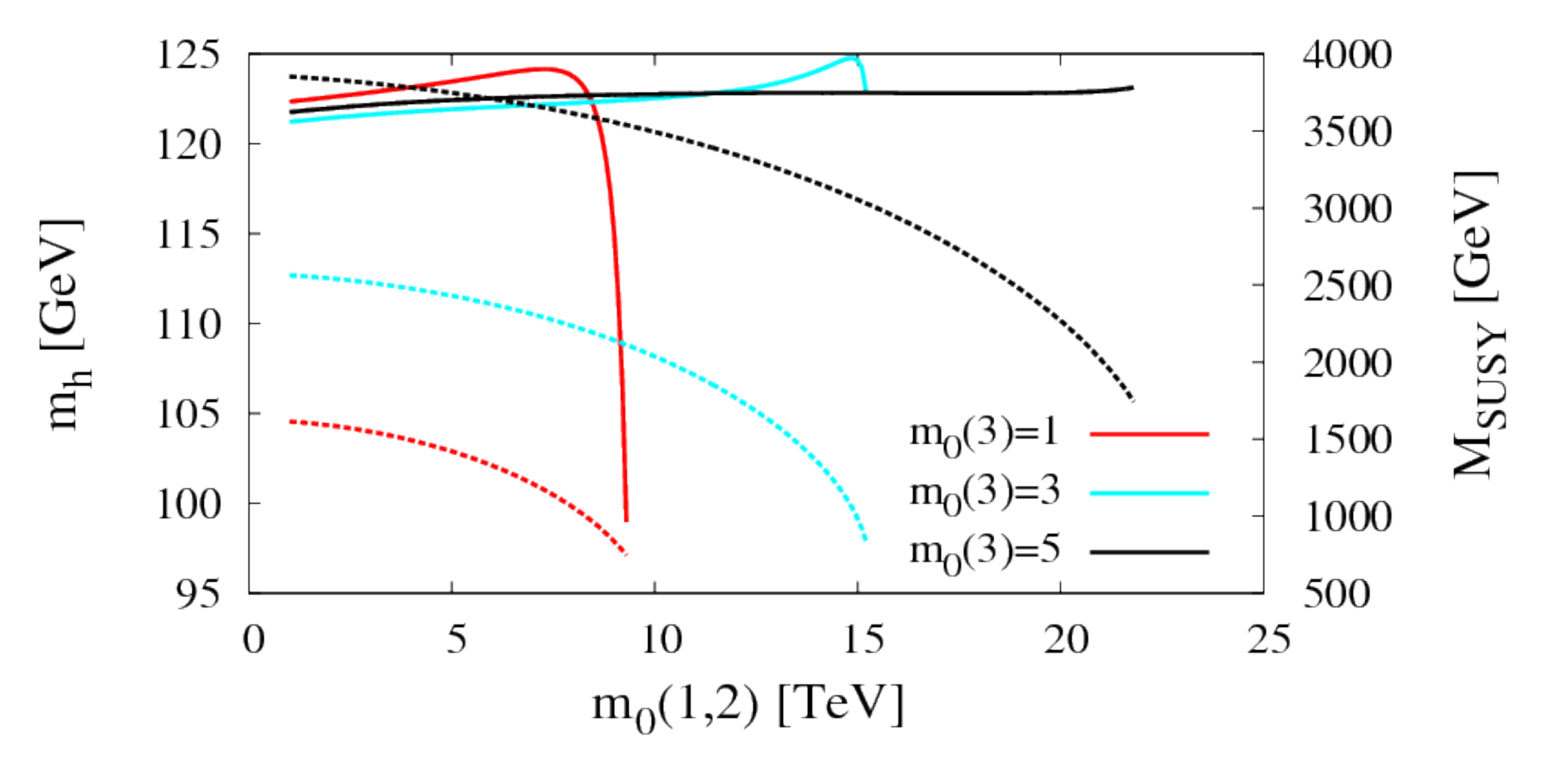}
    \includegraphics[width=\textwidth, height=0.31\textheight]{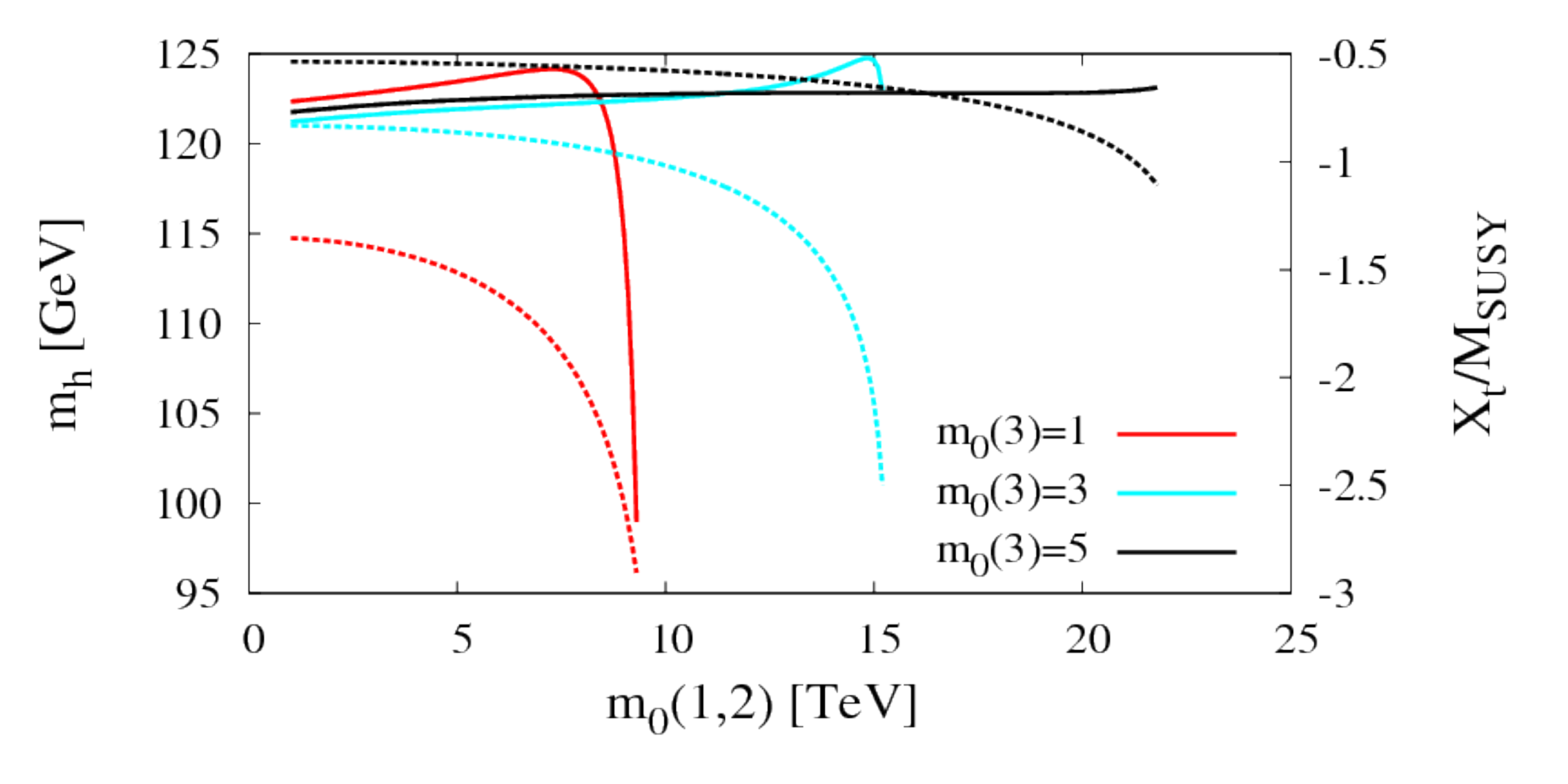}
    \caption{Slices of the part of parameter space shown in Figure \ref{fig:mh_IMH_A0_2000}  for several fixed values of $m_0(3)$ given in TeV on the plots. The solid (dashed) lines correspond to the Higgs mass (the lighter stop mass, $M_{\rm SUSY}$ or $X_t/M_{\rm SUSY}$, respectively).  }
    \label{fig:mh_IMHcuts_A0_2000}
  \end{center}
\end{figure}

In the IH scenario, with $m_0(1,2)\gg m_0(3)$, the RG evolution of $A_t$  can be decoupled from the evolution of
stop masses because the former does not depend on $m_0(1,2)$.   One can enhance $A_t$ by gluino contribution
to the RG running, without enhancing the stop masses, due to the negative 2-loop contribution to the stop masses
from the 1st and 2nd generations. In consequence no large initial values of $A_0$ are needed  for obtaining the Higgs mass in the 125 GeV range,  with  the lighter stop mass in the range 500-1000 GeV and the gluino mass
2-3 TeV.  Thus, $M_{\rm SUSY}\approx 1$ TeV and "maximal" mixing follow quite naturally from the RG evolution.
Moreover, the cancellations in the Higgs potential are significantly smaller
than e.g.\ in the CMSSM focus point.

An example of the predictions for the Higgs boson mass in the IH scenario, for $\tan\beta =10$, $M_{1/2}=1$ TeV
and $A_0=-2$ TeV is shown in Figure \ref{fig:mh_IMH_A0_2000}. 
Very similar predictions are obtained for a range
of $M_{1/2}$ between 0.5 and 2 TeV and $A_0$ between zero and about $-4$ TeV, if
larger $M_{1/2}$ is taken with smaller $|A_0|$.

It can be seen from Figure \ref{fig:mh_IMH_A0_2000}  that  IH scenario
predicts large values of the Higgs boson mass.  In order to better illustrate
the dependence of the Higgs mass across parameter space, particularly near the
tachyonic region, we  present in Figure \ref{fig:mh_IMHcuts_A0_2000} several
slices of the plot from Figure \ref{fig:mh_IMH_A0_2000}  for three fixed
values of $m_0(3)$. For a given value of $m_0(3)$,
the stop masses, as well as $M_{\rm SUSY}$, decrease with increasing
$m_0(1,2)$. In consequence, the 
logarithmic correction to the Higgs mass decreases while the one from
stop mixing increases. Since the latter correction increases polynomially with
$|X_t|/M_{\rm SUSY}$ the Higgs mass initially increases reaching a maximum for
a value of $m_0(1,2)$ corresponding to $|X_t|/M_{\rm SUSY}\approx
2$. Increasing $m_0(1,2)$ further  decreases $m_h$ because the stop-mixing
correction is close to its maximal value and its increase cannot compensate
the decrease of the logarithmic correction. Moreover, for large enough
$|X_t|/M_{\rm SUSY}$ the stop-mixing correction also starts to decrease
resulting in a rapid decrease of the Higgs 
mass.\footnote{A sharp cut-off of some curves  is  partly due to the fact that SOFTSUSY for large $m_0(3)$ is not able to reach very small values of $M_{\rm SUSY}$ before the tachyonic stop is signalized. }

\begin{figure}[t]
  \begin{center}
    \includegraphics[width=\textwidth, height=0.34\textheight]{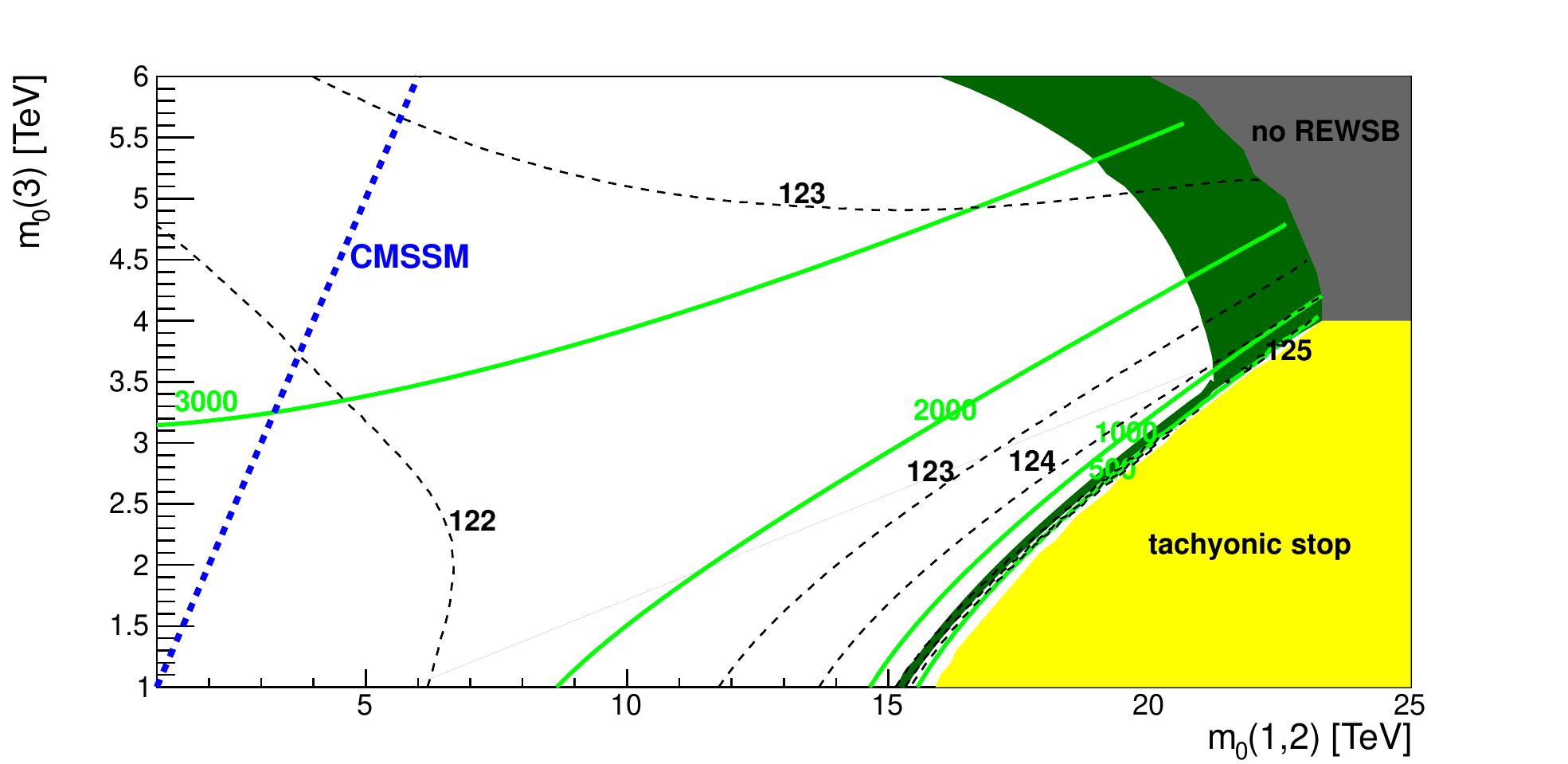}
    \caption{The same as in Figure \ref{fig:mh_IMH_A0_2000} but for $M_{1/2}=1.5$ TeV and $A_0=0$.   }
    \label{fig:mh_IMH_A0_0}
  \end{center}
\end{figure}
\begin{figure}[t]
  \begin{center}
    \includegraphics[width=\textwidth, height=0.31\textheight]{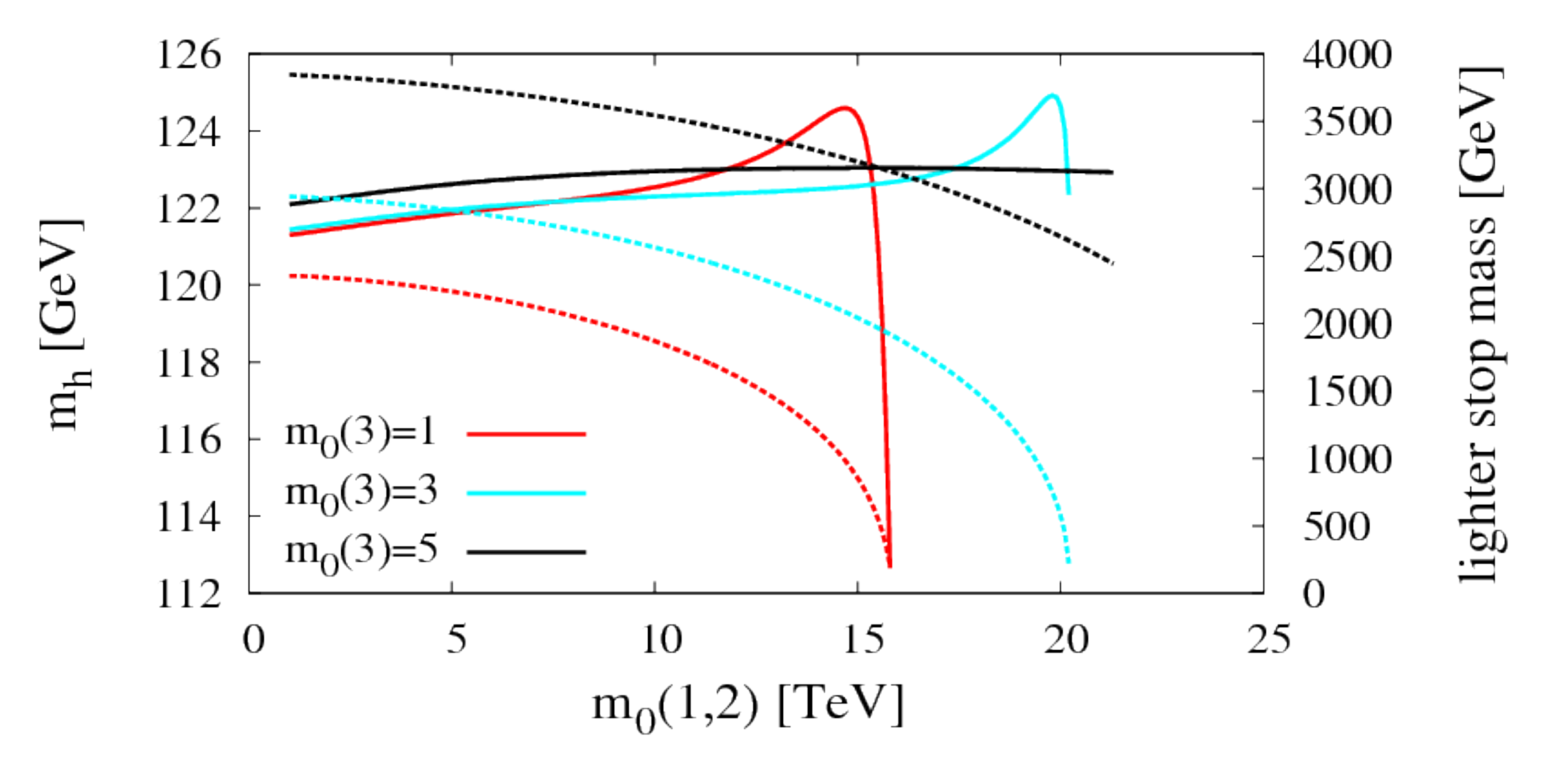}
    \caption{Slices of the part of parameter space shown in Figure \ref{fig:mh_IMH_A0_0}  for several fixed values of $m_0(3)$. The solid (dashed) lines correspond to the Higgs mass (the lighter stop mass).  }
    \label{fig:mh_IMHcuts_A0_0}
  \end{center}
\end{figure}

Non-trivial constraints on the IH scenario follow from the requirement of proper REWSB. The origin of these constraints can be understood by inspecting a dependence of $\mu^2$ on the soft terms defined at the GUT scale which is approximately given by:
\begin{equation}
\label{mucoeff}
 \mu^2\approx -m_{H_u}^2\approx
 1.3M_{1/2}^2+0.1A_0^2-0.35M_{1/2}A_0-0.01m_0(3)^2-0.006m_0(1,2)^2
 \, .
\end{equation}
Notice first the smallness of the coefficient in front of $m_0(3)^2$. Even
though it is negative in the above formula we should emphasize that its sign
depends on the scale at which the coefficients are extracted and to some
extent on the boundary conditions for the soft terms (which influence the GUT
scale values of gauge and Yukawa couplings). It is important to note that the
coefficient in front of $m_0(3)^2$ becomes more and more negative as the Higgs
potential minimization scale, $M_{\rm SUSY}$, increases. On the other hand,
the very small coefficient in front of $m_0(1,2)^2$ is always negative due to
the specific structure of the two-loop RGEs. Even though this coefficient is
very small, large values of $m_0(1,2)$ may drive $\mu^2$ negative. In order to
protect proper REWSB the positive contribution from $M_{1/2}$ and $A_0$ has to
be large enough to ensure that by increasing $m_0(1,2)$ the stops become light
enough to realize ``maximal mixing'' scenario before $\mu^2$ becomes
negative. This effect is seen in Figure \ref{fig:mh_IMH_A0_2000}. The
``maximal mixing''  can be realized only for $m_0(3)$ below some critical
value which in this case is smaller than 5 TeV. For larger values of $m_0(3)$,
$\mu^2$ becomes negative before $m_0(1,2)$ reaches the value corresponding to
$(|X_t|/M_{\rm SUSY})_{\rm max}$ which gives the maximal stop-mixing correction to the Higgs mass.
Constraints from REWSB are also the reason for the lack of maximum in a dependence of the Higgs mass on $m_0(1,2)$ for a large values of $m_0(3)$, as seen in Figure \ref{fig:mh_IMHcuts_A0_2000}.

It is also evident from Figure \ref{fig:mh_IMH_A0_2000}  that the largest Higgs masses correspond to the largest values of $m_0(1,2)$. Therefore, a  heavier Higgs prefers the regions of parameter space where the first and second generation sfermions are decoupled which makes plausible the solution to the FCNC problem. This also can be seen from the opposite perspective. If the solution of the flavor problem relies on the decoupling of the first two generations of sfermions the Higgs mass is expected to be large. In particular, for the parameters used in Figure \ref{fig:mh_IMH_A0_2000}  the requirement of $m_0(1,2)>15$ TeV implies that the lightest Higgs mass has to be not smaller than about 123 GeV. More generally, we found that if the lighter stop mass is at least $\mathcal{O}(500)$ GeV and $m_0(1,2)>15$ TeV the lightest Higgs is necessarily heavier than about 122 GeV. This lower bound on the Higgs mass becomes even more stringent for larger values of $m_0(1,2)$.

Figure \ref{fig:mh_IMH_A0_0}  demonstrates that in the IH scenario the stop-mixing correction to the Higgs boson mass can be maximized even if $A_0=0$. In such a case $A_t$ at the EW scale is generated entirely radiatively by the gaugino masses. This requires gaugino masses somewhat heavier than in the case of non-vanishing $A$-terms. We found that for $A_0=0$, $m_{h}=122$ (125) GeV can be obtained for $M_{1/2}\gtrsim 900$ (1500) GeV.
Notice that in this example the Higgs mass of 124 GeV can be obtained even for
very small values of 
$m_0(3)$.\footnote{One cannot, however, take $m_0(3)$ to be arbitrarily small
  because otherwise the stau would be lighter than the lightest neutralino.  
}
This is because the value of $M_{1/2}=1.5$ TeV is 
large enough to generate stop masses radiatively. 
Smaller values of $m_0(3)$  require  smaller $m_0(1,2)$ to avoid tachyonic stops. Also  Figure \ref{fig:mh_IMH_A0_0}  shows that the requirement of  maximal mixing and of proper REWSB may put an upper bound on $m_0(3)$. Dependence of the Higgs and the lighter stop mass on $m_0(1,2)$ for $A_0=0$ is qualitatively similar to the one with non-vanishing $A$-terms, as seen in Figure \ref{fig:mh_IMHcuts_A0_0}.  

\begin{figure}[t]
  \begin{center}
    \includegraphics[width=\textwidth, height=0.34\textheight]{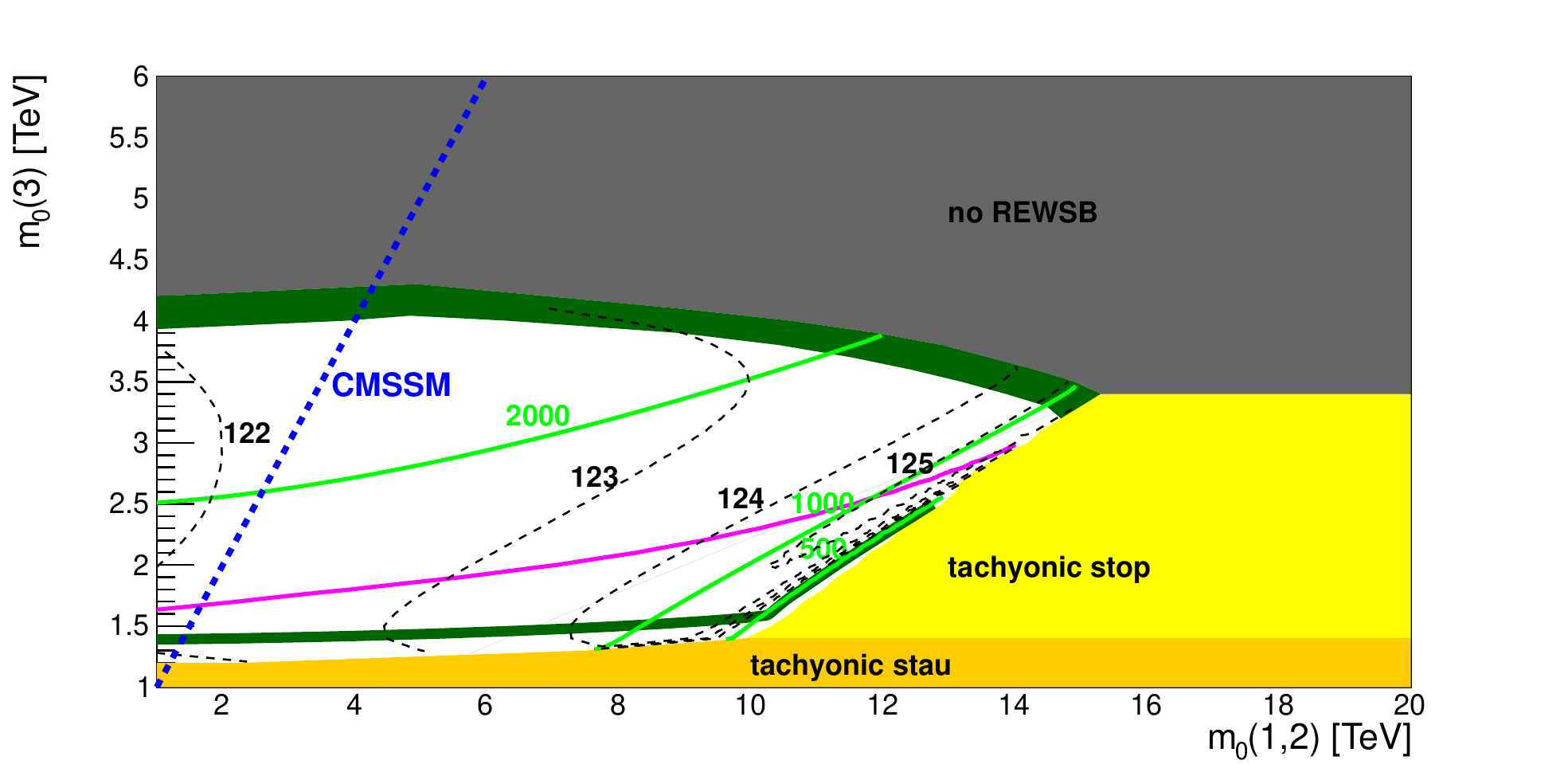}
    \caption{The same as in Figure \ref{fig:mh_IMH_A0_2000} but for
      $\tan\beta=50$ and $m_{H_d}=1.6m_0(3)$. The region below the purple line
      is excluded by BR($B_s\to  \mu^+\mu^-$) at 95\% C.L. The orange region
      is excluded because it predicts a tachyonic stau. }
    \label{fig:mh_IMH_tb50}
  \end{center}
\end{figure}
\begin{figure}[t]
  \begin{center}
    \includegraphics[width=\textwidth, height=0.31\textheight]{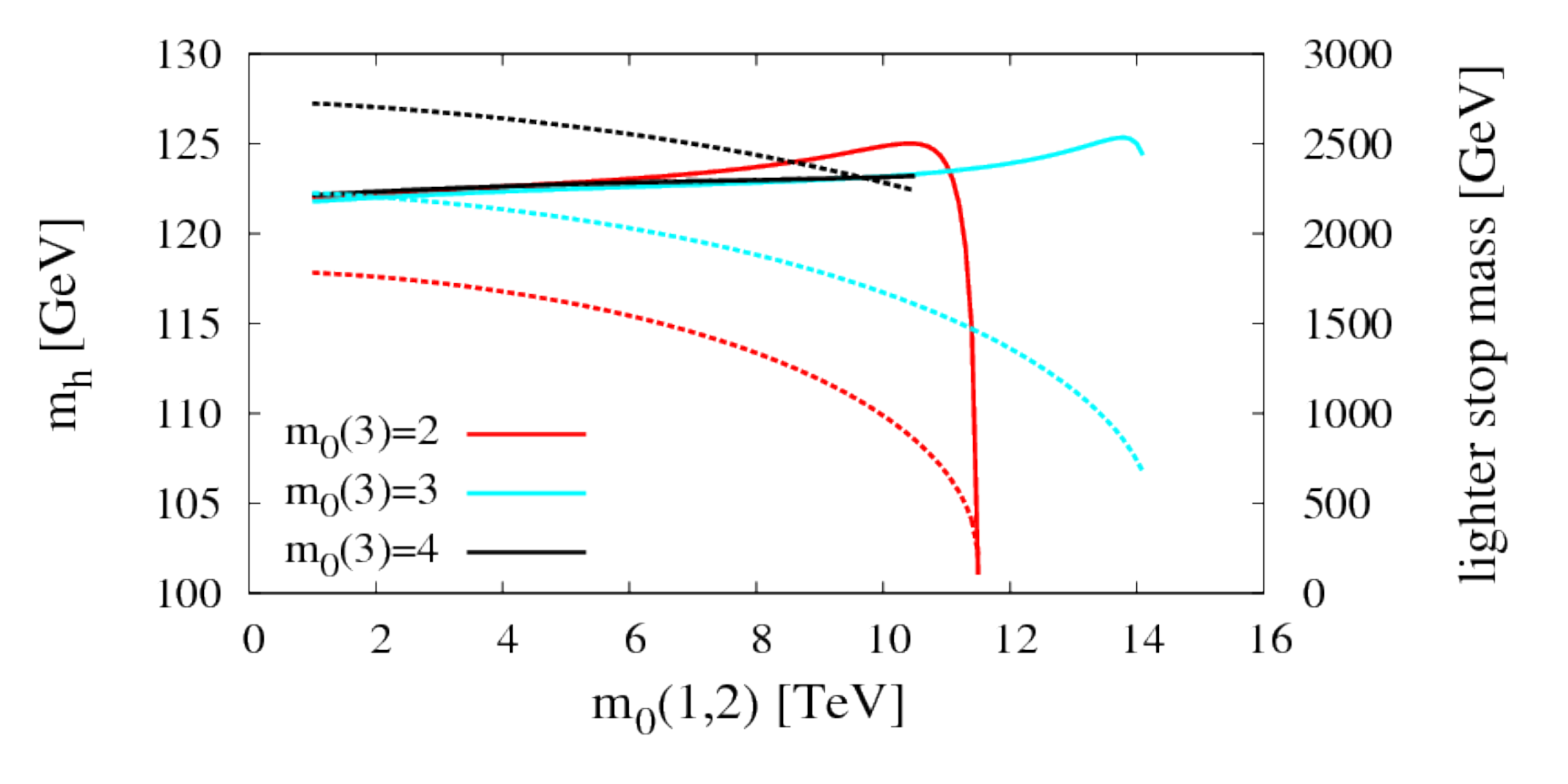}
    \caption{Slices of the part of parameter space shown in Figure \ref{fig:mh_IMH_tb50}  for several fixed values of $m_0(3)$. The solid (dashed) lines correspond to the Higgs mass (the lighter stop mass). }
    \label{fig:mh_IMHcuts_tb50}
  \end{center}
\end{figure}

The IH scenario can be also realized for large values of $\tan\beta$.
In Figure \ref{fig:mh_IMH_tb50} the plot analogous to that from Figure
\ref{fig:mh_IMH_A0_2000} but for $\tan\beta=50$ is presented. In addition, the
splitting between the soft Higgs masses at the GUT scale, $m_0({H_u})=m_0(3)$,
$m_0({H_d})=1.6m_0(3)$, is introduced in order to reduce the impact of the
BR($B_s\to  \mu^+\mu^-$) constraint on the parameter space. 
Even though a dependence of the Higgs and the lighter stop masses on $m_0(1,2)$ for large $\tan\beta$ is qualitatively similar to that for $\tan\beta=10$, as seen in Figure \ref{fig:mh_IMHcuts_tb50}, the upper bounds on $m_0(3)$ and $m_0(1,2)$ set by the condition of positive $\mu^2$ are more
stringent for $\tan\beta=50$. There are two reasons for that. First: the
positive contribution to $\mu^2$ from $M_{1/2}$ and $A_0$ is smaller for
larger values of $\tan\beta$. Second: $m_0({H_d})$ (which is now larger than
$m_0(3)$) gives larger negative contribution to $\mu^2$. Since the tau Yukawa
coupling is now of the same order as the top Yukawa it gives large negative
contribution to the stau masses and some part of parameter space at smaller
values of $m_0(3)$ is excluded because it leads to a tachyonic 
stau.\footnote{As a matter of fact, for $\tan\beta=10$ the tachyonic stau is also present in some part of parameter space but for much smaller values of $m_0(3)$ (outside of the range of the plot in Figure \ref{fig:mh_IMH_A0_2000}).}
However, it can be seen that in spite of these constraints the Higgs boson mass may reach even slightly larger values for $\tan\beta=50$ as compared to the $\tan\beta=10$ case. Moreover, gaugino masses required to obtain a given value of the Higgs mass are somewhat smaller for $\tan\beta=50$. In particular, for $A_0=0$ the Higgs boson mass of 122 (125) GeV can be reached if $M_{1/2}\gtrsim 800$ (1300) GeV.

The examples presented above were designed in such a way that the Higgs mass of 125 GeV is obtained with the smallest possible values of $M_{1/2}$ and $|A_0|$. We should, however, stress that in the IH model larger Higgs masses can be reached if $M_{1/2}$ and/or $|A_0|$ are chosen to be larger. For example, for $M_{1/2}=2.5$ TeV and $A_0=0$ the Higgs mass up to about 127 GeV can be obtained.

One of the reasons why the ``maximal mixing'' generated by IH is interesting is that it requires less fine-tuning than e.g.\ CMSSM to obtain the same Higgs boson mass.
In order to quantify this we define the fine-tuning parameter $\Delta\equiv \max\{\Delta_a\}$, where:\footnote{In the case of large loop corrections to the quartic Higgs coupling, it is more appropriate to refer to $m_h$ rather than $M_Z$ when defining the measure of fine-tuning, see e.g.\ Refs.~\cite{Berezhiani,Papucci}.}
\begin{equation}
 \Delta_a\equiv\left|\frac{\pa \ln m_h}{\pa \ln a}\right| \,.
\end{equation}
The index $a$ stands for any soft term or $\mu$. For a given point in parameter space proper REWSB requires cancellation between the parameters with the precision of order $1/\Delta$.
As explained before, in IH scenario with the ``maximal stop mixing'' the bigger  the value of $M_{1/2}$ is, the smaller value of $A_0$ 
is needed to obtain a given value of the Higgs mass. In other words, the same
Higgs boson mass can be obtained for various values of the ratio 
$A_0/M_{1/2}$.\footnote{Notice that a given value of the Higgs mass can be obtained only for the values of the ratio of $A_0/M_{1/2}$ in some finite range whose size is controlled by the value of $M_{1/2}$. }
The level of fine-tuning strongly depends on the relative values of $A_0$ and
$M_{1/2}$. It can be inferred from the coefficients in eq.~(\ref{mucoeff})
that from the naturalness point of view neither large nor vanishing $A_0$ is the best choice.  We found that for  $\tan\beta =10$ and $A_0=0$ the Higgs mass of 125 GeV requires $\Delta\sim{\mathcal O}(250)$ but for e.g.\ $A_0=-2$ TeV and $M_{1/2}=1$ TeV the Higgs mass of 125 GeV can be reached with $\Delta\sim{\mathcal O}(150)$. Similarly, for $\tan\beta = 50$ the 125 GeV is reached with $\Delta\sim{\mathcal O}(150)$.

It is also interesting to note the existence of a generalized ``focus point''
region. We recall that in the CMSSM the ``focus point'' refers to the region
of parameter space where $m_0\gg M_{1/2}$ \cite{focuspoint,Nath1}. The focus point
in the CMSSM gained a lot of interest since heavy stops required to lift the
Higgs mass above the LEP bound \cite{LEPbound} can be  reconciled with small
contribution of $m_0$ to the Higgs potential and keeping $\mu^2$ close to the
EW scale. Such small values of $\mu^2$ result from the cancellation of the
gaugino contribution to $m_{H_u}^2$ with that of scalars (see
Eq.~(\ref{mucoeff})). In the focus point region of the CMSSM the LEP bound can be satisfied without introducing large fine-tuning of the EW scale. However, in the context of the 125 GeV Higgs the focus point region of the CMSSM is much less attractive because in this part of parameter space the stop mixing is strongly suppressed, unless $A_0$ is large \cite{Feng,Nath2}.
In the IH parameter range, by taking $m_0(1,2)>>m_0(3)>M_{1/2}$, the virtues of the focus point remain but in addition one obtains large stop mixing and the region of 125 GeV for the  Higgs mass is easily reachable.

Our results are in a qualitative agreement with the results of
ref.\ \cite{Baer_NatSUSY}. Somewhat smaller values of $m_h$ reported in
ref.\ \cite{Baer_NatSUSY} are due to the fact that in
ref.\ \cite{Baer_NatSUSY} the computation of the Higgs boson mass is performed
at one-loop level using ISASUGRA program from the ISAJET package
\cite{Isajet}, while we use the 
routines adopted in SOFTSUSY \cite{softsusy} which take into account all
relevant two-loop corrections. 
\footnote{There is also another difference between ISASUGRA and SOFTSUSY. In ISASUGRA the heavy sfermions of the first two generations are decoupled from the RGEs at a scale equal to the heavy-sfermion mass, while SOFTSUSY decouples all the SUSY particles at a common scale $Q=M_Z$. In SOFTSUSY only leading logarithms of the ratios between the heavy-sfermion masses and $M_Z$ are included in the threshold corrections to the gauge and Yukawa couplings. On the other hand, the decoupling procedure adopted in ISASUGRA effectively accounts for the resummation of the logarithms. Simple estimate of the contribution from the non-leading logarithms, which is missing in SOFTSUSY, gives few percent for the sfermion masses of order ${\mathcal O}(10 {\rm TeV})$. }
For example for the benchmark points listed in Table 1 of
ref.\ \cite{Baer_NatSUSY} we find values of $m_h$ larger
by 0.5-2 GeV.
Secondly, in ref.\ \cite{Baer_NatSUSY} the region of light 
(below 1 TeV) average mass 
$\bar{m}_{\tilde{q}}(3) 
= 
\left(m_{\tilde{t}_1}+m_{\tilde{t}_2}+m_{\tilde{b}_1}\right)/3$ is emphasized
as the one that satisfies simple naturalness constraints. Our values of $\bar{m}_{\tilde{q}}(3)$ are 
typically above 1 TeV (because of the masses of the heavier stop and the lighter sbottom) and for the average $\bar{m}_{\tilde{q}}(3)$ in
the range 1-1.5 TeV the upper bound on $m_h$ reported in
ref.\  \cite{Baer_NatSUSY} is 124 GeV, quite close to our result of 125-126
GeV.  We also notice that the fine tuning in our parameter range, although
with heavier third generation spectrum, is similar to that for the benchmark
points of Table 1 of ref.\ \cite{Baer_NatSUSY}, because of our generalized
focus point.


\subsection{Inverted hierarchy models: Abelian flavor symmetries}

So far, we have discussed the predictions for the Higgs boson mass in the IH scenario, with $m_0(1,2)>>m_0(3)$,
without any reference to potential models of such hierarchy. We would like now to recall briefly its link to models of fermion masses based on horizontal symmetries.

Inverted hierarchy was proposed some time ago  \cite{pomarol,ckn} as a way to ease the FCNC and CP constraints
in supersymmetric models. Early ideas did invoke horizontal non-Abelian symmetries for explaining fermion mass hierarchies,  like $U(2)$ \cite{pomarol}, under which first two generations transform as a doublet, whereas the third generation is a singlet. Whereas $U(2)$
models do explain the difference between the first two generations and the third one and therefore can accommodate a hierarchy between the
first two and the third generation of scalars, they do not actually predict it. To our knowledge, the first class of models
in which the inverted hierarchy was really predicted \cite{dgps}
are supersymmetric generalizations of Abelian flavor models of the Froggatt-Nielsen
type \cite{fn}. These models contain an additional Abelian gauge symmetry $U(1)_X$ under which the three fermion generations have different charges (therefore the name horizontal or flavor symmetry), spontaneously broken at a high energy scale by the
vev of (at least) one scalar field $\Phi$, such that $\epsilon = \langle \Phi \rangle / M << 1$, where $M$ is the Planck scale or more
generically the scale where Yukawa couplings are generated.
Quark mass matrices for example, in such models are given, order of magnitude wise, by
\be
h_{ij}^U \ \sim \ \epsilon^{q_i + u_j + h_u} \quad , \quad h_{ij}^D \ \sim \ \epsilon^{q_i + d_j + h_d} \ , \label{abelian1}
\ee
where $q_i$ ($u_i,d_i,h_u,h_d$) denote the $U(1)_X$ charges of the left-handed quarks (right-handed up-quarks, right-handed down-quarks,
$H_u$ and $H_d$, respectively).

A successful fit of the experimental data requires larger charges for the lighter generations
\be
q_1 \ > q_2 \ > q_3 \quad , \quad u_1 \ > u_2 \ > u_3 \quad , \quad d_1 \ > d_2 \ > d_3 \ ,  \label{abelian3}
\ee
one simple example being (see e.g.\cite{cklp})
\be
(q_1,u_1,e_1)  = 3 \ , \ (q_2,u_2,e_2) = 2 \ , \ (q_3,u_3,e_3) = 0 \ , \ (d_1,l_1) = 1 \ , \ (d_2,l_2) = 0 \ , \ (d_3,l_3) = 0 \ .  \label{abelian4}
\ee
Scalar soft masses in Abelian flavor models are typically of the form (only the diagonal in flavour entries are relevant for our present discussion)
\be
m_{i}^2 \ = \ \widetilde{Q}_i \langle D \rangle \ + \ (m_{i}^F)^2 \ , \label{abelian5}
\ee
where $\widetilde{Q}_i \langle D \rangle$ are D-term contributions for the scalar of charge $\widetilde{Q}_i$, whereas  $(m_{i}^F)^2$ are F-term contributions.
D-term contributions were argued to be naturally generated (at least) in
effective string models \cite{bddp}, to be positive and, in certain
circumstances, to be dominant over the F-term contributions. It is then clear
from (\ref{abelian3}), (\ref{abelian4}) and (\ref{abelian5})  that  the hierarchy of the masses of scalars  is {\it predicted to be inverted} with respect to the hierarchy of fermion masses.

Whereas Abelian models naturally predict the inverted hierarchy,
they do not generically predict approximate degeneracy among the first two generations, unlike their non-Abelian cousins.
This leads to possible tension with FCNC constraints, which have to be analyzed in some details (for previous works see e.g.\ \cite{dgps,cklp}).
We would like also to point out that recently there were various other explicit realizations of the inverted hierarchy, based on geometric localization and non-Abelian family models \cite{nonabelian}. 

From the perspective of our present paper, 
the  inverted hierarchy models do generically predict also a splitting between the first-two-generation sfermions as well as intragenerational splitting
\be
m_{Q_i} \not= m_{U_i} \not= m_{D_i} \ .  \label{abelian6}
\ee
A relevant question is therefore what changes are expected in the results presented so far by relaxing the hypothesis of degenerate first two generations and of
$m_{Q_i}= m_{U_i}= m_{D_i}$. Since the first two generations are very heavy, we could expect  large RGE effects. The RGEs
of all scalar soft masses and in particular of the third generation of squarks and of the Higgs scalars depend  at 1-loop level on
the combination (see e.g. \cite{martin})
\be
S \ = \ {\rm Tr} (Y m^2) \ = \ m_{H_u}^2-m_{H_d}^2+\sum_{i=1}^3[m_{Q_i}^2-2m_{U_i}^2+m_{D_i}^2-m_{L_i}^2+m_{E_i}^2]  \ ,  \label{abelian7}
\ee
where the trace is over the whole spectrum of MSSM states and which,  under our assumption until now,  is zero at high-energy. Interestingly enough, in Abelian flavor models with D-term dominance of the type put forward
in this section, the quantity S is equal to
\be
S \ = \ {\rm Tr} (Y \widetilde{Q}) \  \langle D \rangle \ .  \label{abelian8}
\ee
However, ${\rm Tr} (Y \widetilde{Q})$ has to vanish (or to be very small) for phenomenological reasons, as argued in various papers \cite{nelson} and therefore our  conclusions concerning the running of soft terms and the fine-tuning remain unchanged. In particular, for the charge assignment (\ref{abelian4}) $S=0$ and the prediction for the Higgs boson mass are almost identical to those presented in the previous plots with degenerate first two generations.

If the $U(1)_X$ charges of fermions are not chosen in such a way that $S$ vanishes then $S$ is generically of order $\langle D \rangle$ i.e. of the same order as the first two generation soft masses.
Since $S$ enters RGEs at the one-loop level its effect on the RG running of soft scalar masses (at least for non-colored scalars) would generically be much larger than the two-loop effect from heavy first two generations. The contribution from $S$ to the EW scale soft scalar masses is determined by the hypercharge assignment and is approximately given by:
\begin{equation}
 m^2_f = -0.05 Y_f S \,,
\end{equation}
where $Y_f$ is the hypercharge of fermion $f$. In particular, $S$ contributes to the EW scale value of $m_{H_u}^2\approx-0.025 S$ (compare with Eq.~(\ref{mucoeff})) so for the values of $\sqrt{S}\sim{\mathcal O}(10 {\rm TeV})$  it is the dominant contribution implying that proper REWSB is possible only if $S$ is positive. For $S>0$, $S$ gives positive contribution to $m_U^2\approx0.035 S$ and compensates the negative two-loop effect (see coefficients in Eq.~(\ref{mUcoeff})) which is crucial for obtaining the maximal stop mixing. The contribution to $m_Q^2\approx-0.008 S$ is negative but relatively small. The largest contribution from $S$ is received by $m_E^2\approx-0.05S$. Since this contribution is negative one can expect that large values of $S$ typically result in tachyonic staus. Therefore, we conclude that in IH models the maximal stop mixing consistent with REWSB is possible only if $|S|$ is smaller than the $D$-term contribution to the first and second generation squarks.

Let us also briefly comment on the effect of the mass splitting within the third generation of sfermions. Such splitting may be present due to arbitrary ${\mathcal{O}}(1)$ coefficients of the diagonal F-term in eq.~(\ref{abelian5}). The individual contributions from the scalars to $\mu^2$ is approximately given by:
\begin{eqnarray}
 \mu^2\approx
&&\!\!\!\!\!\!\!\!
-0.6\,m_0({H_u})^2+0.35\,m_0({Q_3})^2+0.25\,m_0({U_3})^2
\nn&&\!\!\!\!\!\!\!\!
-0.025\,m_0({H_d})^2+0.025\,m_0({D_3})^2-0.025\,m_0({L_3})^2+0.025\,m_0({E_3})^2 \,.
\end{eqnarray}
For universal soft scalar masses, the overall contribution from the scalars to $\mu^2$ is quite insensitive to the scalar mass mainly because the contributions from $m_0({H_u})$, $m_0({Q_3})$ and $m_0({U_3})$ approximately cancel out. Therefore, it is rather clear that splitting between these three scalar masses may substantially change the overall picture. If $m_0({H_u})\gtrsim \sqrt{0.6m_0({Q_3})^2+0.4m_0({U_3})^2 }$ at the GUT scale, the scalars give negative contribution to $\mu^2$ which makes the upper bound on $m_0(1,2)$ from the REWSB constraint more stringent unless gauginos are heavier. In the opposite case, the inverted hierarchy and the maximal stop mixing may be realized for lighter gauginos than in the case of degeneracy between $m_0({H_u})$ and universal third generation scalar masses.


\section{More phenomenology}
\label{sec:pheno}

\begin{table}[t!]\vspace{1.5cm}
\centering
\begin{tabular}{|c|cccc|}
\hline
\hline
                 & Point A & Point B & Point C & Point D \\
\hline
$M_{1/2}$         & 1000  & 1500  & 1500 & 1000  \\
$m_0(3)$        & 3700 & 3400  & 3800 &  3300   \\
$m_0(1,2)$         & 17690 & 21070  & 22500 & 14500  \\
$A_0$         & -2000 & 0  & 0 &  -2000  \\
$m_0({H_d})/m_0(3)$ & 1  &  1  &  1 &  1.6   \\
$\tan\beta$      & 10  & 10 & 10 &  50   \\

\hline
$\mu$            & 888  & 698  & 452 & 457  \\

\hline
$m_h$            & 125  & 125 & 125.1 & 125.3  \\
$m_H$            & 3541  & 3154   & 3477 & 3487   \\
$m_A$            & 3541 & 3154  & 3477 & 3487  \\
$m_{H^{\pm}}$    & 3542  & 3155    & 3478 & 3488  \\

\hline
$m_{\tilde{\chi}^0_{1,2}}$
                 & 444, 813 & 647, 707  & 448, 461 & 419, 467  \\
$m_{\tilde{\chi}^0_{3,4}}$
                 & 891, 940 & 722, 1284  & 677, 1286 & 483, 869   \\

$m_{\tilde{\chi}^{\pm}_{1,2}}$
                 & 812, 940 & 700, 1284  & 455, 1286 & 457, 869  \\
$m_{\tilde{g}}$  & 2465 & 3530  & 3545 & 2432   \\

\hline $m_{ \tilde{u}_{L,R}}$
                 & 17675, 17675  & 21116, 21119  & 22526, 22532 & 14531, 14510   \\
$m_{\tilde{t}_{1,2}}$
                 & 476, 1801 & 699, 1581  & 505, 1632 & 979, 1274  \\
\hline $m_{ \tilde{d}_{L,R}}$
                 & 17675, 17685  & 21116, 21120 & 22526, 22533 & 14531, 14541   \\
$m_{\tilde{b}_{1,2}}$
                 & 1784, 2926  & 1555, 2717  & 1610, 2933 & 1176, 1584  \\
\hline
$m_{\tilde{\nu}_{1,2}}$
                 & 17680  & 21068  & 22495 & 14481 \\
$m_{\tilde{\nu}_{3}}$
                 & 3466 & 3107  & 3480 & 2367  \\
\hline
$m_{ \tilde{e}_{L,R}}$
                & 17681, 17686  & 21069, 21068  & 22495, 22497 & 14482, 14528   \\
$m_{\tilde{\tau}_{1,2}}$
                & 3467, 3580 & 3108, 3257  & 3481, 3645 & 1853, 2368   \\

\hline

$\Omega_{DM}h^{2}$ &  0.111 & 0.118  & 0.021 & 0.116  \\
BR$(b\to s\gamma)$ &  $2.92\times 10^{-4} $  &  $2.89\times 10^{-4} $ & $2.66\times 10^{-4} $ &	$1\times 10^{-4} $ 	\\
BR$(B_s\to \mu^+ \mu^-)$ &  $3.07\times 10^{-9} $ & $3.07\times 10^{-9} $ & $3.07\times 10^{-9} $ & $3.61\times 10^{-9}$ 	\\
$a_{\mu}^{\rm SUSY}$  & $1\times 10^{-12} $ & $7\times 10^{-13} $  & $5\times 10^{-13} $ &  $1\times 10^{-11} $   \\
\hline

\hline
\hline
\end{tabular}
\caption{Several benchmark points with the inverted scalar mass hierarchy characterized by large stop-mixing contribution to the Higgs mass. Point A is an example of small mass splitting between the bino LSP and the lighter stop. Points B and D have a mixed higgsino-bino LSP. Point C is characterized by mainly Higgsino LSP.}
\label{tab:benchmarks}
\end{table}

\subsection{Dark matter}

In the IH scenario there are two distinctive ways to make the relic abundance of the LSP\footnote{We use MicrOMEGAs \cite{Micromega} to compute the relic abundance of the LSP, as well as BR$(b\to s \gamma)$, BR($B_s\to  \mu^+\mu^-$) and SUSY contribution to muon anomalous magnetic moment, $a_{\mu}^{\rm SUSY}$. }
compatible with the cosmological observations:
\begin{itemize}
 \item Stop-coannihilation for values of $m_0(1,2)$ close to the boundary of
   the region where stops are tachyonic. The region of stop-coannihilation is
   generically present in the IH scenario for some intermediate range of
   $m_0(1,2)$ where the splitting between the stop NLSP and the neutralino LSP
   masses is small. $\Omega_{\rm DM}h^2$ consistent with the WMAP bound is
   obtained for very small range of $m_0(1,2)$ and typically requires an
   adjustment of $m_0(1,2)$ with a precision of order $10^{-3}$. This is
   because bino is the LSP which usually leads to too large values of
   $\Omega_{\rm DM}h^2$ \cite{welltempered}. It is interesting to note that
   the stop-coannihilation region often coincides with the region where the
   stop-mixing correction to the Higgs mass is maximized. A benchmark point
   illustrating such a case is given as a Point A in Table
   \ref{tab:benchmarks}.

 \item Higgsino LSP or mixed Higgsino-bino LSP for values of $m_0(1,2)$ close
   to the boundary of the region where $\mu^2<0$. This resembles the focus
   point scenario in the  CMSSM where for appropriately large values of $m_0$,
   $|\mu|$ can become smaller than $M_1$. In the present case small values of
   $|\mu|$ are obtained not only due to large values of $m_0(3)$ but also because
   of even larger values of $m_0(1,2)$. If the Higgsino is the LSP its relic
   abundance usually turns out to be too small but if the LSP is a mixed state
   of bino and Higgsino the  $2 \sigma$ WMAP bound can be    accommodated
   \cite{welltempered}.  The  region  with  a significant component of
   Higgsino in the LSP is much larger than the stop-coannihilation region and
   does not require very precise choice of parameters. However, the large stop
   mixing correction to the Higgs mass in the region with the Higgsino LSP is
   present only in some part of it where the Higgsino and the lighter stop are
   light simultaneously. In our plots the Higgsino and the lighter stop are
   both light in the part of the dark green region in the vicinity of the
   border between the grey and the yellow regions (corresponding to $\mu^2<0$
   and tachyonic stop, respectively). The corresponding benchmark points are
   given as Points B and D in Table \ref{tab:benchmarks} for $\tan\beta=10$
   and $\tan\beta=50$, respectively. 

\end{itemize}


\subsection{$b\to s \gamma$}

The SM prediction \cite{Misiak} for BR$(b\to s \gamma)$ is about 1$\sigma$ below the experimental central value \cite{HFAG}:
\begin{align}
&{\rm BR}^{\rm SM}(b\to s \gamma)
= \left(3.15\pm0.23\right)\times 10^{-4}
\,,
\nn[4pt]
&{\rm BR}^{\rm exp}(b\to s \gamma)
=\left(3.55\pm0.24\pm0.09\right)\times
10^{-4}
\,.
\end{align}
In the MSSM there are two important contributions to BR$(b\to s \gamma)$ in addition to the SM ones: the charged Higgs contribution and the chargino-squark contribution \cite{DeGaGi}. The former one always increases the SM model prediction.  In the IH scenario for $\tan\beta=10$ the charged Higgs is far above TeV so this contribution is negligible. On the other hand, the chargino contribution is negative (with respect to the SM) for $\mu M_3>0$  because the sign of this contribution is the same as ${\rm sgn} \left(\mu A_t \right)$
\footnote{There is also part of the chargino contribution which has the same sign as ${\rm sgn} \left(-\mu M_2 \right)$ so strictly speaking the chargino contribution may be negative also in some part of the parameter space where $\mu A_t>0$. A detailed discussion on the sign of the chargino contribution can be found e.g.\ in \cite{bop}. }
and maximal stop mixing in IH scenario can be obtained only for $A_t<0$. It is
also important to add that the main chargino contribution is maximized for the
maximal  stop mixing and grows linearly with $\tan\beta$. In the region of
parameter space with the maximal correction to the Higgs mass from stop mixing
BR$(b\to s \gamma)$  is typically between $1\sigma$ and 3$\sigma$ below the experimental central value for $\tan\beta=10$. The largest deviation from the experimental central value occurs for the points with the lightest Higgsino.

Since the chargino contribution is proportional to $\tan\beta$, BR$(b\to s \gamma)$ is even more restrictive for $\tan\beta=50$. In this case BR$(b\to s \gamma)$ in the region of parameter space with the ``maximal mixing'' is far below the experimental value. For the benchmark point D in Table \ref{tab:benchmarks} with the Higgsino-bino LSP the discrepancy between the theory and experiment is about $8\sigma$. This discrepancy can be relaxed to some extent if the Higgsino is heavier but even for $\mu\approx1$ TeV, BR$(b\to s \gamma)$ is about $5\sigma$ below the experimental value.

\subsection{$B_s\to  \mu^+\mu^-$}

The LHCb  upper limit for BR($B_s\to  \mu^+\mu^-$) is \cite{Bsmumu_LHCb}:
\begin{equation}
  {\rm BR}(B_s\to  \mu^+\mu^-)<4.5\times 10^{-9} {\rm at}\ 95\% {\rm C.L.}  \,,
\end{equation}
which is now very close to the SM prediction \cite{Bsmumu_SM}:
\begin{equation}
  {\rm BR}^{\rm SM}(B_s\to  \mu^+\mu^-)=\left(3.2\pm0.2\right)\times 10^{-9} \,.
\end{equation}
This leaves very little room  for contributions from the new physics. In the MSSM, BR($B_s\to  \mu^+\mu^-$)  probes the region of large $\tan\beta$ since the dominant MSSM contribution is proportional to $A_t \tan^6\beta/m_A^4$ \cite{Bsmumu_MSSM}. At large $\tan\beta$, the CP-odd Higgs mass 
is given by:
\begin{equation}
  m_A^2\approx m_{H_d}^2-m_{H_u}^2-M_Z^2 \,,
\end{equation}
where $m_{H_d}$ and $m_{H_u}$ should be understood as the soft masses at the low scale.
The RG running of $m_{H_d}^2-m_{H_u}^2$ at one-loop level is given by:
\begin{equation}
\label{mHdmHuRGE}
 8\pi^2\frac{\de}{\de t}(m_{H_d}^2-m_{H_u}^2)=3h_t^2\widetilde{X}_t-3h_b^2\widetilde{X}_b-h_{\tau}^2\widetilde{X}_{\tau}+\frac{3}{5}g_1^2S \,,
\end{equation}
where
\begin{align}
 &\widetilde{X}_t=m_{Q_3}^2+m_{U_3}^2+m_{H_u}^2+A_t^2\,, \\
 &\widetilde{X}_b=m_{Q_3}^2+m_{D_3}^2+m_{H_d}^2+A_b^2\,, \\
 &\widetilde{X}_{\tau}=m_{L_3}^2+m_{E_3}^2+m_{H_d}^2+A_{\tau}^2\,,
\end{align}
$S$ is given by Eq.~(\ref{abelian7}) and
$t\equiv \ln(M_{\rm in}/Q)$.  We omitted terms proportional to the first and second generation Yukawa couplings which are negligible. Since the bottom and tau Yukawa couplings give the negative contribution to the RG running, the pseudoscalar Higgs becomes very light or even tachyonic when the top, bottom and tau Yukawa couplings are of the same order. In particular for $\tan\beta=50$ (and the other input parameters as specified below Eq.~(\ref{Atcoeff})) $m_{H_d}^2-m_{H_u}^2$ at the scale $Q=1.5$ TeV is given by:
\begin{equation}
\label{mHdmHucoeff}
 m_{H_d}^2-m_{H_u}^2\approx 0.6M_{1/2}^2-0.02A_0M_{1/2}+0.03A_0^2+0.2m_0(3)^2-0.005m_0(1,2)^2 \,.
\end{equation}
The corresponding Yukawa couplings at the GUT scale equal: $h_t=0.56$, $h_b=0.35$, $h_{\tau}=0.51$. From the above formula it is clear that the CP-odd Higgs mass is driven to smaller values when $m_0(1,2)$ increases.

BR($B_s\to \mu^+\mu^-$) can easily be brought to phenomenologically acceptable values by splitting the soft Higgs masses at the GUT scale in such a way that $m_0({H_d})>m_0({H_u})$ since such splitting makes the pseudoscalar Higgs heavier.

\subsection{MSSM spectrum and LHC phenomenology}

In the IH scenario with ``the maximal mixing'' the lighter stop is expected to
be the lightest colored sparticle, in the range from 500 GeV to 1 TeV. The magnitude of the two loop effect of the heavy first two generations on the stau masses is controlled by the weak gauge coupling so they are typically heavier than the stops. Masses of the sfermions of the first and second generations are set in the first approximation by $m_0(1,2)$ so they are much heavier than sfermions of the third generation.

Gluino  masses are in the range $2-3$ TeV and other gauginos are correspondingly lighter, for universal $M_{1/2}$.
As explained earlier,  higgsino mass may be in the range 200-400 GeV.

If gluino is much heavier than the lighter stop,  the stop pair production
dominates SUSY production cross-section at the LHC. In such a case the limits
on the stop mass are much weaker because the stop pair production
cross-section is several orders of magnitude smaller than that of the gluinos
with the same mass. The LHC limits on the direct stop production have not been
presented so far.  It is argued in \cite{Papucci} that the existing LHC
searches with jets and missing energy may have already excluded
$m_{\tilde{t}}\lesssim 300$ GeV in some particular cases when the Higgsino is
the LSP and a main decay mode is $\tilde{t}\rightarrow b
\tilde{\chi}^{\pm}$. However, this limit is highly model dependent and
e.g.\ no lower mass limit for right-handed stop was found in \cite{Papucci}
if bino is the LSP. Moreover, as stated before no official limits on direct
stop pair production from the LHC experiments are available 
yet.\footnote{
In \cite{Atlas_gmsb_directstop} the ATLAS constraints on the direct stop pair production have been presented in a particular simplified model (motivated by gauge mediated SUSY breaking) with a gravitino LSP but the stop decay chain in such a model differs very much from those typical for the IH scenario.}

For large $\tan\beta$ the spectrum of the third generation is more compressed. In such a case the stop mass splitting is smaller so the lighter stop mass is expected to be a bit larger than for moderate $\tan\beta$ - not much below 1 TeV (in order to get $M_{\rm SUSY}\approx 1$ TeV as preferred by the 125 GeV Higgs mass), see benchmark point D in Table \ref{tab:benchmarks}. In addition, sbottoms are expected to be in the TeV range. Constraining direct production of sbottoms is experimentally less challenging than that of stops so in IH scenario at large $\tan\beta$ sbottoms may be discovered (or ruled out) before the stops. Nevertheless, for the time being sbottom masses above about 400 GeV are consistent with the experiment \cite{sbottom_direct}.

From the above discussion it should be clear that the IH scenario is very weakly constrained at the moment.

\section{Conclusions}
\label{sec:concl}

The idea that the first two generations of sfermions are much heavier than the third one has been promoted in the past as a way to ease the supersymmetric FCNC problem,  without violating the naturalness principle. Also, the sfermion mass spectrum can then be linked to the fermion masses in models based on horizontal symmetries. The early LHC results, putting stronger lower limits on the masses of the first-two-generation squarks than on stops and sbottoms, add some attractiveness to this idea. In this paper we have shown that this scenario  predicts large stop mixing as a consequence of the RG evolution, with vanishing or small A-terms at the high scale.  For the lightest stop mass to be at least $\mathcal{ O}(0.5)$ TeV and assuming proper REWSB the Higgs boson is necessarily heavy, easily in the range 122-127 GeV. In particular, if the masses of the first-two-generation sfermions are above 15 TeV, the above conditions place a lower bound on the Higgs boson mass of about 122 GeV and the bound becomes more stringent as the masses of the first-two-generation sfermions increase. The Higgs boson mass of 125 GeV requires the universal gaugino mass to be about 1.5 TeV for vanishing $A$-terms at the GUT scale. It  can be substantially smaller if negative $A_0$ is assumed, e.g.\ for $A_0=-2$ TeV it is enough to have $M_{1/2}\approx 1$ TeV.
This scenario is only moderately fine-tuned.  The LSP remains an interesting dark matter candidate, particularly
when it has a strong higgsino component or the stop NLSP is degenerate with the LSP to the extent which allows for efficient stop-coannihilations.  The parameter range considered in this paper looks like a good bet
for the MSSM.

\section*{Acknowledgments}

This work has been partially supported by STFC. We would like to thank G. Gersdorff and R. Ziegler for useful discussions on inverted
hierarchy models. MB would like to thank B. C. Allanach for helpful discussions, as well as T. Hahn, S. Heinemeyer and P. Slavich for useful correspondence. The present research was supported in part by the ERC Advanced Investigator Grant no. 226371 ``Mass Hierarchy and Particle Physics at the TeV Scale'' (MassTeV), by the contract PITN-GA-2009-237920 and by the French ANR TAPDMS ANR-09-JCJC-0146. S.P. acknowledges partial support by the contract PITN-GA-2009-237920 UNILHC and by the National Science Centre in Poland under research grant DEC-2011/01/M/ST2/02466.


\end{document}